\documentclass[11pt,a4paper]{article}
\pdfoutput=1
\usepackage{jheppub,yfonts}

\usepackage{amsmath}
\usepackage{amssymb}
\usepackage{amsthm}
\usepackage{mathrsfs}
\usepackage{graphicx}
\usepackage{fancyhdr}
\usepackage{array}
\usepackage{latexsym}
\usepackage[all]{xy}
\usepackage{eufrak}
\usepackage{euscript}
\usepackage{enumerate}
\usepackage{dsfont}
\usepackage{slashed}
\usepackage{hyperref}
\usepackage{caption}

\graphicspath{{figures//}}

\newcommand{\be}{\begin{equation}}
\newcommand{\ee}{\end{equation}}
\newcommand{\bea}{\begin{eqnarray}}
\newcommand{\eea}{\end{eqnarray}}

\title{Hydrodynamic transport coefficients for the non-conformal quark-gluon plasma from holography}

\author[a]{Stefano I.~Finazzo,}
\author[a]{Romulo Rougemont,}
\author[a]{Hugo Marrochio,}
\author[a,b]{Jorge Noronha}

\affiliation[a]{Instituto de F\'{i}sica, Universidade de S\~{a}o Paulo, C.P. 66318, 05315-970, S\~{a}o Paulo, SP, Brazil}
\affiliation[b]{Department of Physics, Columbia University, 538 West 120th Street, New York, NY 10027, USA}

\date{\today}

\abstract{In this paper we obtain holographic formulas for the transport coefficients $\kappa$ and $\tau_\pi$ present in the second-order derivative expansion of relativistic hydrodynamics in curved spacetime associated with a non-conformal strongly coupled plasma described holographically by an Einstein+Scalar action in the bulk. We compute these coefficients as functions of the temperature in a bottom-up non-conformal model that is tuned to reproduce lattice QCD thermodynamics at zero baryon chemical potential. We directly compute, besides the speed of sound, 6 other transport coefficients that appear at second-order in the derivative expansion. We also give an estimate for the temperature dependence of 11 other transport coefficients taking into account the simplest contribution from non-conformal effects that appear near the QCD crossover phase transition. Using these results, we construct an Israel-Stewart-like theory in flat spacetime containing 13 of these 17 transport coefficients that should be suitable for phenomenological applications in the context of numerical hydrodynamic simulations of the strongly-coupled, non-conformal quark-gluon plasma. Using several different approximations, we give parametrizations for the temperature dependence of all the second-order transport coefficients that appear in this theory in a format that can be easily implemented in existing numerical hydrodynamic codes.}
   

\emailAdd{stefano@if.usp.br, romulo@if.usp.br, hcmarroc@if.usp.br, noronha@if.usp.br}
 
\begin{document}

\maketitle

\setlength{\parskip}{8pt}


\section{Introduction}
\label{introduction}

After the discovery of the quark-gluon plasma (QGP) in ultra-relativistic heavy ion collisions \cite{qgp}, a lot of effort has been put towards understanding how the spatial anisotropies present in the initial state are converted into the final flow of hadrons. Relativistic dissipative hydrodynamics has played an important role in our current view of the complicated spacetime evolution of the QGP formed in heavy ion collisions (for a recent review see \cite{Heinz:2013th}). The overall picture that is consistent with experimental data is that before hadronization the QGP evolves in time and space as a relativistic fluid with minimal dissipative effects. Indeed, current estimates \cite{Heinz:2013th} for the shear viscosity to entropy density ratio, $\eta/s$, of the QGP obtained by comparison to data are in the ballpark of the very small value $\eta/s=1/(4\pi)$ \cite{Kovtun:2004de} found in a broad class of strongly-coupled non-Abelian plasmas described by the gauge/gravity duality \cite{adscft1,adscft2,adscft3}. This suggests that the gauge/gravity duality may be useful for the study of the non-equilibrium properties of strongly interacting plasmas that are similar (if not quantitatively at least qualitatively so) to the QGP and, thus, several applications have been studied over the last years (see, for instance, the review \cite{CasalderreySolana:2011us}). 

In fact, we shall show in this paper that a simple bottom-up holographic model that is able to describe (some of) the thermodynamic properties of the QGP\footnote{As it will be clear later in the paper, this model cannot describe aspects of QCD that are directly related to chiral symmetry.} near the crossover phase transition \cite{Aoki:2006we} can be instrumental in providing estimates for the temperature dependence of a large number of second order transport coefficients that appear in consistent theories of (non-conformal) dissipative relativistic hydrodynamics.

This paper is organized as follows. In the next section we shall review the dissipative hydrodynamic theory obtained at second-order in gradients in the case of a non-conformal, relativistic plasma (in the absence of conserved charges such as the baryon number) in a curved spacetime \cite{romatschke}. At this order in the gradient expansion, there are 17 coefficients (besides the speed of sound) that may possess some nontrivial temperature dependence (especially near the phase transition)\footnote{We note that many more coefficients would be needed in the case where spatial isotropy in the equilibrium state is lost, as it occurs in anisotropic hydrodynamics \cite{Martinez:2010sc,Florkowski:2010cf,Ryblewski:2012rr,Martinez:2012tu,Florkowski:2013lza,Florkowski:2013lya,Strickland:2014pga} and in fluids in the presence of strong magnetic fields (see, for instance, \cite{Huang:2011dc}). However, these interesting generalizations will not be pursued here.}. In Section \ref{sec2} we present our method to holographically compute the second order transport coefficients $\kappa$ and $\tau_\pi$ in strongly-coupled plasmas that are described by a bulk action including the metric and a dynamical scalar field (see also Appendix \ref{apa}). In Section \ref{sec4} we give the details of the bottom-up holographic model we use and fix its parameters through a comparison to lattice QCD thermodynamics. In Section \ref{transportsec} we compute the temperature dependence of several transport coefficients for this holographic bottom-up model. In Section \ref{ISsection} we use the 2nd-order gradient theory defined in Section \ref{2ndorderhydrosection} to write an Israel-Stewart-like hydrodynamic theory in flat spacetime with 13 transport coefficients that could be implemented in numerical hydrodynamics. Also, a guide to the temperature dependence of the several 2nd-order transport coefficients considered in this paper (given in terms of fitting functions that could be easily used in numerical hydrodynamics) can be found in Appendix \ref{summaryhydro}. Furthermore, in Appendix \ref{appendixhugo} we perform a linear stability analysis around the static equilibrium for the non-conformal, 2nd order gradient expansion theory discussed in Section \ref{2ndorderhydrosection}. Our conclusions and outlook can be found in Section \ref{concsec}.

The reader that is mostly interested in the hydrodynamic discussions and the specific temperature dependence of the transport coefficients (shown in Figs.\ \ref{fig:kappa} to \ref{fig:taupibulk}) may want to focus on Sections \ref{2ndorderhydrosection}, \ref{ISsection}, and Appendices \ref{summaryhydro} and \ref{appendixhugo}. The other sections are devoted to more detailed calculations involving the gauge/gravity duality.

Throughout this paper we use natural units $c=\hbar=k_B=1$ and a {\it mostly plus} metric signature. Also, we use capital Latin indices to denote the bulk coordinates $x^M=(t,x,y,z,u)$ while Greek indices $x^\mu = (t,x,y,z)$ denote 4-dimensional coordinates.

\section{Second-order non-conformal hydrodynamics via the gradient expansion}
\label{2ndorderhydrosection}

Relativistic dissipative hydrodynamics can be viewed as a type of effective theory for the long wavelength, low frequency behavior of an interacting system at finite temperature and/or chemical potential \cite{landau,forster}. Such an effective theory may be constructed at weak coupling in the case of a dilute gas \cite{chapman,LandauKine,grad,burnett,degroot,Denicol:2012cn,bookkremer} whose microscopic behavior can be described by a Boltzmann-like equation for the system's effective quasi-particles \cite{Jeon:1994if,Jeon:1995zm}. On the other hand, at strong coupling the fluid-gravity correspondence \cite{Bhattacharyya:2008jc} provides an adequate framework to study the effects of spacetime gradients in a strongly coupled fluid. 

In general, dissipation is included directly at the level of the equations of motion\footnote{For recent discussions including attempts to formulate dissipative hydrodynamics in terms of an effective action see, for instance, \cite{Torrieri:2011ne,Dubovsky:2011sj,Kovtun:2014hpa}.}, which in the absence of conserved charges, correspond solely to the conservation of energy and momentum 
\be
\nabla_\mu T^{\mu\nu} = 0\,,
\ee
where $\nabla_\mu$ is the covariant spacetime derivative in a curved 4-dimensional spacetime described by a metric $g_{\mu\nu}$, and $T^{\mu\nu}$ is the expectation value of the system's energy-momentum tensor operator. We shall consider here matter described by a relativistic quantum field theory giving the equation of state, $P=P(\varepsilon)$, where $\varepsilon$ and $P$ are the local energy density and pressure of the fluid, respectively. The equation of state gives rise to the speed of sound in the fluid, $c_s=\sqrt{dP/d\varepsilon}$. The basic idea of the gradient expansion is that the macroscopic degrees of freedom in the long wavelength, low frequency limit are only the local energy density, $\varepsilon$, 4-velocity, $u_\mu$, metric, $g_{\mu\nu}$ (in curved spacetime), and their gradients. In fact, the energy-momentum tensor can be generically decomposed as
\be
T^{\mu\nu} = \varepsilon\, u^\mu u^\nu + P \Delta^{\mu\nu}+\pi^{\mu\nu}+\Delta^{\mu\nu}\Pi
\label{defineTmunuhydro}
\ee
where the flow obeys $u_\mu u^\mu=-1$, and $\Delta^{\mu\nu}=g^{\mu\nu}+u^\mu u^\nu$ is a local projection operator transverse to the flow. Note that such a decomposition inherently assumes that there is a well defined local rest frame (for examples of quantum field theories in far from equilibrium conditions without a local rest frame see Ref.\ \cite{Arnold:2014jva}). Dissipation generally appears due to a nonzero shear stress tensor
\be
\pi^{\mu\nu} = \Delta^{\mu\nu\alpha\beta}T_{\alpha\beta}\,,
\ee 
which is transverse to the flow\footnote{Note we use the Landau frame, i.e., $u_\mu T^{\mu\nu}=-\varepsilon \,u^\nu$ \cite{landau}.}, symmetric, and traceless due to the definition of the tensor projector
\be
\Delta^{\mu\nu\alpha\beta} = \frac{1}{2}\left(\Delta^{\mu\alpha}\Delta^{\nu\beta}+\Delta^{\mu\beta}\Delta^{\nu\alpha}\right)-\frac{1}{3}\Delta^{\mu\nu}\Delta^{\alpha\beta}\,.
\ee 
The last term in (\ref{defineTmunuhydro}) denotes the dissipative contribution to the energy-momentum tensor with non-vanishing trace, $\Pi$, called the bulk viscous pressure. In terms of (\ref{defineTmunuhydro}), one can show that the conservation of energy and momentum become
\bea
D\varepsilon+(\varepsilon+P+\Pi)\theta +\frac{1}{2}\pi_{\mu\nu}\sigma^{\mu\nu} &=& 0\,, \nonumber \\
(\varepsilon+P+\Pi)Du^\mu+\nabla_\perp^\mu (P+\Pi) +\Delta^\mu_\nu\nabla_\alpha \pi^{\alpha\nu} &=& 0\,,
\label{conservationeqs}
\eea
where $D=u^\mu \nabla_\mu$ is the comoving derivative, $\nabla_\perp^\alpha=\Delta^{\alpha\beta}\nabla_\beta$ is the derivative transverse to the flow, $\theta = \nabla_\mu u^\mu$ is the scalar expansion rate, and $\sigma_{\mu\nu}=2\Delta_{\mu\nu}^{\alpha\beta}\nabla_\alpha u_\beta$ is the shear tensor. The energy conservation equation can be written in terms of the equilibrium entropy density, $s=(\varepsilon+P)/T$, as follows
\be
\nabla_\mu (s u^\mu)=D s + s\theta = -\frac{\pi_{\mu\nu}\sigma^{\mu\nu}}{2T}-\frac{\Pi \theta}{T}\,.
\label{entropyproduction}
\ee

In the gradient expansion approach, since only $\varepsilon$ and $u_\mu$ are the hydrodynamical variables, the dissipative components $\pi_{\mu\nu}$ and $\Pi$ must be expressed solely in terms of derivatives of these quantities. To first order in gradients, this can be easily done and one finds
\be
\pi^{\mu\nu}=-\eta \sigma^{\mu\nu}\,,\qquad  \Pi = -\zeta \theta\,,
\label{NS}
\ee
where $\eta$ is the shear viscosity and $\zeta$ is the bulk viscosity, respectively. Note that in this case the second law of thermodynamics in (\ref{entropyproduction}) imposes that $\eta,\zeta \geq 0$. If one uses the expressions above for the dissipative contributions in $T^{\mu\nu}$, the conservation equations represent the relativistic extension of the well-known Navier-Stokes (NS) equations \cite{landau} \footnote{Another way to understand how dissipation appears is to notice that, for instance, in this NS fluid the inclusion of $\pi_{\mu\nu}$ breaks the time reversal invariance present in the ideal fluid equations of motion. However, it is possible to find nontrivial fluid patterns involving second order gradients where $\pi_{\mu\nu}$ is nonzero but time reversal is not broken - see \cite{Hatta:2014gqa,Hatta:2014gga}.}.

In kinetic theory, the transport coefficients $\eta$ and $\zeta$ are proportional to their corresponding mean free paths, $\ell$ \footnote{Note that the mean free path for bulk viscosity is different than that for shear viscosity \cite{Arnold:2006fz}. However, for simplicity, we shall denote any mean free path here by $\ell$.}. One can now see how the power counting scheme adopted in the gradient expansion works. Since $\ell$ is a microscopic scale and $\varepsilon$ and $u_\mu$ are taken to be slowly varying functions of time and space, one can associate with their gradients a characteristic (macroscopic) length scale $\sim 1/L_{macro}$ such that $\ell/L_{macro} \ll 1$. Therefore, terms such as those in (\ref{NS}) are taken to be of order 1 in the so-called Knudsen number $K_n \sim \ell/L_{macro}$ \footnote{We remark that the Knudsen ``number" is actually a field since it depends on the spacetime coordinates. Moreover, in general one may consider several types of Knudsen numbers associated with different properties of the flow, see for instance \cite{Niemi:2014wta}.}. Clearly, the continuous description of the system as a fluid hinges on the assumption that $K_n$ is sufficiently small. However, given that dissipation only appears at order 1 in this expansion, one may also entertain the case in which $K_n$ is still sufficiently small to ensure a well-defined continuous description but the flow is such that higher order terms may be taken into account. Nevertheless, one should keep in mind that the radius of convergence of the gradient series is limited by the first nonzero non-hydrodynamical quasinormal mode, as recently shown in \cite{Heller:2013fn} in the context of strongly coupled gauge theories and discussed earlier in \cite{Denicol:2011fa} in the context of kinetic theory.

For instance, in the early stages of a ultrarelativistic nucleus-nucleus collision \cite{Schenke:2012wb}, the local energy density and flow are expected to have sizable gradients and corrections of second order in $K_n$ may be relevant at that stage of the QGP evolution. Also, as emphasized in \cite{Niemi:2014wta}, in collisions involving smaller systems such as proton-nucleus collisions at the LHC, the need for higher order Knudsen number corrections may be even more pressing. Therefore, it is reasonable to ask what are the expressions for $\pi_{\mu\nu}$ and $\Pi$ including $\mathcal{O}(K_n^2)$ terms. Generalizing the previous analysis involving 2nd order terms in a conformal fluid done in \cite{brsss}, Romatschke proposed in \cite{romatschke} the following expansion for the dissipative parts of a non-conformal relativistic fluid in curved spacetime valid at $\mathcal{O}(K_n^2)$ 
\bea
\pi^{\mu\nu} &=& -\eta \sigma^{\mu\nu} + \eta \tau_\pi \left(D\sigma^{\langle \mu\nu\rangle}+\frac{\theta}{3}\sigma^{\mu\nu}\right) + \kappa \left(\mathcal{R}^{\langle \mu\nu\rangle}-2u_\alpha u_\beta \mathcal{R}^{\alpha \langle \mu\nu \rangle \beta}\right) \nonumber \\ &+& \lambda_1  \sigma_\lambda^{\langle \mu} \sigma^{\nu\rangle\lambda}+ \lambda_2 \sigma_\lambda^{\langle \mu} \Omega^{\nu\rangle\lambda} -\lambda_3  \Omega_\lambda^{\langle \mu} \Omega^{\nu\rangle\lambda} \nonumber \\ &+&2 \kappa^{*} \,u_\alpha u_\beta \mathcal{R}^{\alpha \langle \mu\nu \rangle \beta} + \eta \tau_\pi^{*}\, \sigma^{\mu\nu}\,\frac{\theta}{3}+\lambda_4 \nabla^{\langle \mu}\ln s \,\nabla^{\nu\rangle} \ln s\,,
\label{defineshearstress}
\eea
and
\bea
\Pi  &=& -\zeta \theta + \zeta \tau_\Pi \,D\theta +\xi_1 \sigma_{\mu\nu}\sigma^{\mu\nu} +\xi_2\, \theta^2 \nonumber \\ &+& \xi_3 \Omega_{\mu\nu}\Omega^{\mu\nu}+\xi_4 \nabla_\mu^{\perp} \ln s\,\nabla_\perp^\mu \ln s+\xi_5 \mathcal{R} + \xi_6 u^\mu u^\nu \mathcal{R}_{\mu\nu}\,,
\label{definebulk}
\eea
where $\mathcal{R}^\lambda_{\mu\sigma\nu}$ is the Riemann tensor, $\mathcal{R}_{\mu\nu} = \mathcal{R}^\lambda_{\mu\nu\lambda}$ is the Ricci tensor, and $\mathcal{R} = g_{\mu\nu}\mathcal{R}^{\mu\nu}$ is the Ricci scalar \cite{weinberg}. Moreover, we have also defined the vorticity tensor $\Omega_{\mu\nu} = \frac{1}{2}\left(\nabla_\mu^\perp u_\nu - \nabla_\nu^\perp u_\mu\right)$ and the usual notation $B^{\langle \mu\nu\rangle}=\Delta^{\mu\nu}_{\alpha\beta}B^{\alpha\beta}$ for the traceless, symmetric, and transverse part of a second rank tensor $B^{\mu\nu}$. Eqs.\ (\ref{defineshearstress}) and (\ref{definebulk}) can then be used in the conservation equations (\ref{conservationeqs}) to define the equations of motion for a non-conformal fluid in a curved spacetime valid at second order in gradients\footnote{Due to the fact that spacetime covariant derivatives generally do not commute in a curved spacetime, even when the metric is not dynamical (though still nontrivial) quantities such as $\mathcal{R}^\lambda_{\mu\nu\alpha}$, $\mathcal{R}_{\mu\nu}$, and $\mathcal{R}$ are expected to appear in the equations of motion for $\varepsilon$ and $u_\mu$. Also, we note that the expressions in (\ref{defineshearstress}) and (\ref{definebulk}) are in agreement with the corresponding terms (in flat spacetime) at $\mathcal{O}(K_n^2)$ found in \cite{Denicol:2012cn}.}. 

One can see that besides the speed of sound squared $c_s^2 = dP/d\varepsilon$ that is already present in ideal hydrodynamics and the coefficients $\eta$ and $\zeta$ that appeared at 1st order, there are now altogether 15 new transport coefficients that appear at second order in the gradient expansion. Following \cite{moore}, one may distinguish these coefficients by separating out those that are of {\it thermodynamical} origin and those that are not. The set of coefficients $\kappa$, $\kappa^{*}$, $\lambda_3$, $\lambda_4$, $\xi_3$, $\xi_4$, $\xi_5$, $\xi_6$ can be determined via Kubo formulas involving only equilibrium quantities and Euclidean two- and three-point functions of the energy-momentum tensor components and, thus, they are of thermodynamical origin being suitable for lattice calculations \cite{moore}. However, the other coefficients $\eta$, $\zeta$, $\tau_\pi$, $\tau_\pi^{*}$, $\tau_\Pi$, $\lambda_1$, $\lambda_2$, $\xi_1$, and $\xi_2$ are associated with quantities that define the dissipative properties of the theory (for instance, $\eta$ is proportional to the imaginary part of a retarded Green's function) such as $\sigma_{\mu\nu}$ and are, thus, of dynamical origin \cite{moore}.

There are several interesting points regarding these second order terms. First, as discussed in \cite{romatschke,hydro1,hydro2,hydro3,moore} not all of these coefficients are independent and we shall get back to this point in Section \ref{transportsec} where we compute several of these coefficients in the holographic model defined in Section \ref{sec4}. Second, for a conformal plasma only the terms that transform homogeneously under a Weyl transformation\footnote{Under a Weyl transformation, $g_{\mu\nu} \to e^{-2\mathcal{\omega}} g_{\mu\nu}$, where $\mathcal{\omega}$ is an arbitrary (positive-definite) scalar function. In a conformal plasma, $T^{\mu\nu} \to e^{6 \mathcal{\omega}} T^{\mu\nu}$ while the temperature transforms as $T \to e^{\mathcal{\omega}}T$, see \cite{brsss}.} are present \cite{brsss} and this implies that $\kappa^{*}$, $\tau_{\pi^{*}}$, $\lambda_4$, $\zeta$, $\tau_\Pi$, $\xi_{1,2,3,4,5,6}$ vanish in the conformal limit (therefore, in this case, $\Pi=0$). Third, even though $\kappa$, $\kappa^{*}$, $\xi_5$, and $\xi_6$ do not contribute to the equations of motion in flat spacetime, they do contribute to the Kubo formulas for the other coefficients relevant to flat spacetime hydrodynamics and should then be taken into account, as discussed in \cite{brsss,moore}. Moreover, in a non-conformal fluid all of these coefficients may be nontrivial functions of the temperature, i.e., $\eta=\eta(T)$, especially near a phase transition (even if of the crossover type). This shows how challenging numerical 2nd order hydrodynamics can be if all of these temperature dependent transport coefficients are taken into account. Clearly, as long as these 2nd order gradient terms can be taken as small corrections, their effect should be under control. However, it is not clear at the moment if this is indeed the case for the type of event-by-event hydrodynamic simulations fed by the complicated initial conditions that describe the early stages of a heavy ion collision \cite{Schenke:2012wb}.

Furthermore, unfortunately the equations of motion defined by (\ref{defineshearstress}) and (\ref{definebulk}) cannot be directly implemented in numerical hydrodynamic codes because, for instance, they are linearly unstable against small fluctuations around the static equilibrium (see Appendix \ref{appendixhugo}). In fact, in Section \ref{ISsection} we propose another second order theory that is more suitable for numerical investigations using the current relativistic hydrodynamic codes. This theory can be considered as a type of ``UV completion" of the 2nd order theory in (\ref{defineshearstress}) and (\ref{definebulk}) in the sense that it possesses a well defined (and causal) UV behavior (at least in the linear regime) but its long wavelength, low frequency asymptotic hydrodynamical solution is identical to the one obtained by (\ref{defineshearstress}) and (\ref{definebulk}) with the same transport coefficients. We shall discuss these points in detail in Section \ref{ISsection}.

\section{Holographic calculation of the 2nd order coefficients $\kappa$ and $\tau_\pi$}
\label{sec2}

In this section we discuss how we are going to evaluate the 2nd order hydrodynamic transport coefficients $\kappa$ and $\tau_\pi$, originally defined in \cite{brsss}, using the gauge/gravity duality \cite{adscft1,adscft2,adscft3}. First, we shall consider the case where the action in the bulk corresponds to 5-dimensional pure gravity (with a negative cosmological constant) and then later we will generalize this discussion for the case where the bulk action contains a dynamical scalar field.  

\subsection{Renormalized pure gravity action}

In order to compute these transport coefficients, it is sufficient to consider only the small frequency $\omega$, small momentum $q$ limit of the $(xy,xy)$-component of the retarded propagator of the stress-energy tensor of the boundary quantum field theory, which can be written\footnote{This retarded Green's function is given by $G_R^{xy,xy}(\omega,\vec{q})=-i \int_{\mathbb{R}^{1,3}} d^4x\, e^{i (\omega t - \vec{q}\cdot \vec{x})}\,\theta(t)\langle [\hat{T}^{xy}(t,\vec{x}),\hat{T}^{xy}(0,\vec{0})  ]\rangle$.} in momentum space as follows \cite{romatschke}
\begin{align}
G_R^{xy,xy}(\omega,q)=P-i\eta\omega+\left(\eta\tau_\pi-\frac{\kappa}{2}+\kappa^*\right)\omega^2-\frac{\kappa}{2}q^2 +\mathcal{O}(\omega q^2,\omega^3),
\label{2.1}
\end{align}
where $\kappa^*$ is a second order hydrodynamic coefficient which is non-vanishing only for non-conformal fluids (see Eq.\ (\ref{defineshearstress})), being related to $\kappa$ by the following constraint \cite{hydro1,hydro2,hydro3,moore}
\begin{align}
\kappa^*=\kappa-\frac{T}{2}\frac{d\kappa}{dT}\,.
\label{2.2}
\end{align}

According to the gauge/gravity duality dictionary, the stress-energy tensor of the plasma is sourced in the partition function of the boundary quantum field theory by the boundary value of a classical metric perturbation placed over an asymptotically AdS bulk. In (\ref{2.1}), it was assumed that the $xy$-component of the metric perturbation has no dependence on the $x$- and $y$-directions, such that the 4-momentum of its Fourier mode is given by $k_\mu=(-\omega,q_x,q_y,q_z)=(-\omega,0,0,q)$. From (\ref{2.1}) and (\ref{2.2}), one obtains the following Kubo formulas for $\kappa$ and $\tau_\pi$ (valid for both conformal and non-conformal fluids)\footnote{We thank G.~Moore and K.~Sohrabi for discussions about these coefficients.}
\begin{align}
\kappa&=-\lim_{q\rightarrow 0}\lim_{\omega\rightarrow 0}\frac{\partial^2G_R^{xy,xy}(\omega,q)}{\partial q^2}, \label{2.3}\\
\tau_\pi&=\frac{1}{2\eta} \left(\lim_{q\rightarrow 0}\lim_{\omega\rightarrow 0}\frac{\partial^2G_R^{xy,xy}(\omega,q)}{\partial \omega^2}-\kappa+T\frac{d\kappa}{dT}\right).\label{2.4}
\end{align}

In order to compute $\kappa$ and $\tau_\pi$ from (\ref{2.3}) and (\ref{2.4}) via holography, we need to evaluate the renormalized on-shell bulk action and extract from it the retarded graviton propagator by following the prescription proposed in \cite{ss}, which was later justified and generalized in \cite{prescription-justification-1,prescription-justification-2,full-prescription}. In the case of pure gravity, the regularized action is defined by the sum of the Einstein-Hilbert (EH) action, the Gibbons-Hawking-York (GHY) action\footnote{The GHY action needs to be added to the EH action in order to properly define the variational problem with Dirichlet boundary conditions for the metric tensor when the spacetime manifold has a boundary.} \cite{ghy1,ghy2} and the counterterm action\footnote{The counterterm action is obtained through the holographic renormalization procedure \cite{ren1,ren2,ren3,ren4,ren5}.} \cite{brsss,pss}
\begin{align}
S_{\textrm{reg}}&=S_{\textrm{EH}}+S_{\textrm{GHY}}+S_{\textrm{CT}}\nonumber\\
&=\frac{1}{16\pi G_5}\left\{\int_{\mathcal{M}_5}d^5x\sqrt{-g}\left[R(g)-2\Lambda\right] +2\int_{\partial\mathcal{M}_5}d^4x\sqrt{-\gamma}K(\gamma)+\right.\nonumber\\
&\left.\,\,\,\,\,\,\,\,\,\,\,\,\,\,\,\,\,\,\,\,\,\,\,\,\,\,\,\,\,\,\, -\frac{6}{L}\int_{\partial\mathcal{M}_5}d^4x\sqrt{-\gamma} \left[1+\frac{L^2}{2}\mathcal{P}-\frac{L^4}{12}\left(\mathcal{P}^{\mu\nu}\mathcal{P}_{\mu\nu} -\mathcal{P}^2\right)\ln(\epsilon)\right]\right\},
\label{2.5}
\end{align}
where $G_5$ is the five dimensional Newton's constant and the cosmological constant enforcing the existence of asymptotically AdS$_5$ geometries with radius $L$ as solutions of Einstein's equations is given by $\Lambda=-6/L^2$. The metric induced at the boundary is given by $\gamma_{MN}=g_{MN}-\hat{n}_M\hat{n}_N$, where $\hat{n}_M$ is an outward directed unit vector orthogonal to the boundary. In a coordinate chart where the boundary of the asymptotically AdS space is at the value $u=0$ of the radial coordinate, this is given by $\hat{n}_M=-\delta_M^u\sqrt{g_{uu}}\Rightarrow \hat{n}^M=-\delta^M_u/\sqrt{g_{uu}}$, and the metric induced at the regularizing boundary surface\footnote{Strictly speaking, this quantity diverges at $u=0$. Therefore, in order to regularize quantities of interest, we introduce an ultraviolet cutoff $\epsilon\ll 1$ for the radial coordinate near the boundary, which must be taken to zero at the very end of the calculations after all the divergent terms in the on-shell gravity action have been canceled.} $u=\epsilon$ can be simply written as
\begin{align}
\gamma_{\mu\nu}=g_{\mu\nu}\biggr|_{u=\epsilon}.
\label{2.6}
\end{align}
The extrinsic curvature of the boundary of an asymptotically AdS space is given by (see \cite{mcgreevy} for a review)
\begin{align}
K(\gamma)=\frac{\hat{n}^u}{2}\gamma^{\mu\nu}\partial_u\gamma_{\mu\nu}=-\frac{\sqrt{g^{uu}}}{2}\gamma^{\mu\nu}\gamma '_{\mu\nu},
\label{2.7}
\end{align}
where the prime denotes a derivative in the radial direction. The boundary sectional curvature tensor and the boundary sectional curvature scalar are defined, respectively, by \cite{ren4}
\begin{align}
\mathcal{P}_{\mu\nu}=\frac{1}{2}\left(\mathcal{R}_{\mu\nu}(\gamma)-\frac{1}{6}\mathcal{R}(\gamma)\gamma_{\mu\nu}\right),\,\,\, \mathcal{P}=\gamma^{\mu\nu}\mathcal{P}_{\mu\nu},
\label{2.8}
\end{align}
where $\mathcal{R}_{\mu\nu}(\gamma)$ and $\mathcal{R}(\gamma)$ are the Ricci tensor and the Ricci scalar evaluated using the induced metric on the boundary surface.

Let us now consider a small perturbation in the metric, $h_{MN}(u,t,z)$, placed over a diagonal and isotropic gravitational background\footnote{The index $(0)$ refers to the undisturbed background.}, $g_{MN}^{(0)}(u)$, which is assumed to be asymptotically AdS. Since we are only interested in the $(xy,xy)$-component of the retarded propagator of the boundary stress-energy tensor, as discussed in \cite{pss}, if one fixes the gauge defined by the subsidiary condition $h_{M u}=0$, one only needs to consider $h_{xy}(u,t,z)\neq 0$ and set to zero all the other components of the metric perturbation since the linearized equation of motion for the $xy$-perturbation decouples from the other components of the metric perturbation in this gauge. Therefore, we consider the following disturbed line element
\begin{align}
ds^2&=g_{MN}(u,t,z)dx^M dx^N\nonumber\\
&=g_{MN}^{(0)}(u)dx^M dx^N+2h_{xy}(u,t,z)dxdy\nonumber\\
&=g_{uu}(u)du^2-g_{tt}(u)dt^2+g_{xx}(u)(dx^2+dy^2+dz^2)+2g_{xx}(u)\phi(u,t,z)dxdy,
\label{2.9}
\end{align}
where we defined $\phi(u,t,z)=h^x_y(u,t,z)$ and, with the sign convention used in (\ref{2.9}), we are considering the $tt$-component of the metric to be $-g_{tt}(u)$, with $g_{tt}(u)>0$.

Let us now explicitly show that the EH action for the disturbed metric, up to second order in the metric perturbation $\phi$, can be written as the action for a massless scalar field in the undisturbed background $g_{MN}^{(0)}$ plus total derivatives which shall contribute to the final expression for the regularized on-shell boundary gravity action\footnote{We thank R.~Critelli for discussions concerning this derivation.}. Up to $\mathcal{O}(\phi^2)$, we find
\begin{align}
\sqrt{-g}R(g)\approx \left[\sqrt{-g^{(0)}}\left(1-\frac{\phi^2}{2}\right)\right]\,\left[R^{(0)}+\frac{3}{2} g^{MN}_{(0)}\partial_M\phi\partial_N\phi+2\phi\nabla_{(0)}^2\phi+\frac{g'_{xx}}{g_{uu}g_{xx}}\phi\phi '\right].
\label{2.10}
\end{align}
Since we are assuming that the background metric is a solution of Einstein's equations, we can make use of these equations to write down two useful relations for our purposes. The first one comes from contracting the background metric with its equation of motion
\begin{align}
g^{MN}_{(0)}\left[R_{MN}^{(0)}-\frac{g_{MN}^{(0)}}{2}\left(R^{(0)}-2\Lambda\right)\right]=0\Rightarrow R^{(0)}=\frac{2D}{D-2}\Lambda=-\frac{20}{L^2}\,.
\label{2.11}
\end{align}
The second useful relation comes from substituting (\ref{2.11}) into the $xx$-component of the Einstein's equations for the background metric
\begin{align}
R_{xx}^{(0)}-\frac{g_{xx}}{2}\left(-\frac{8}{L^2}\right)=0\Rightarrow \frac{4}{L^2}=-\frac{1}{g_{xx}}\left[ -\frac{\nabla_{(0)}^2 g_{xx}}{2} + \frac{g_{(0)}^{MN}\partial_M g_{xx}\partial_N g_{xx}}{2 g_{xx}} \right].
\label{2.12}
\end{align}
Substituting (\ref{2.11}) into (\ref{2.10}), plugging the result into the EH action and integrating by parts the term proportional to $\phi\nabla_{(0)}^2\phi$, we obtain, up to $\mathcal{O}(\phi^2)$
\begin{align}
S_{\textrm{EH}}&\approx -\frac{1}{2\pi L^2G_5}V_4\int_\epsilon^{u_H}du\sqrt{-g^{(0)}}+\frac{1}{16\pi G_5} \int_{\mathcal{M}_5}d^5x\sqrt{-g^{(0)}}\left[-\frac{1}{2}g_{(0)}^{MN}\partial_M\phi\partial_N\phi+\mathcal{I}\right] +\nonumber\\
&+\frac{1}{8\pi G_5}\int_{\partial\mathcal{M}_5}d^4x\sqrt{-\gamma^{(0)}g^{uu}}\phi\phi ' \biggr|_\epsilon^{u_H},
\label{2.13}
\end{align}
where $V_4=\int_{\partial\mathcal{M}_5}d^4x$ is the 4-volume of the boundary, $u=u_H$ is the position of the background black hole horizon in the radial coordinate\footnote{If the background has no event horizon (or some kind of infrared wall), then one must take $u_H\rightarrow\infty$.} and
\begin{align}
\int_{\mathcal{M}_5}d^5x\sqrt{-g^{(0)}}\mathcal{I}&=\int_{\mathcal{M}_5}d^5x\sqrt{-g^{(0)}}\left[\frac{g'_{xx}}{g_{uu}g_{xx}} \phi\phi '+\frac{4}{L^2}\phi^2\right]\nonumber\\
&=\int_{\mathcal{M}_5}d^5x\sqrt{-g^{(0)}}\left[\frac{g_{(0)}^{MN}\partial_M g_{xx}\partial_N(\phi^2)}{2g_{xx}} +\frac{\nabla_{(0)}^2 g_{xx}}{2g_{xx}}\phi^2 - \frac{g_{(0)}^{MN}\partial_M g_{xx}\partial_N g_{xx}}{2 g_{xx}^2}\phi^2 \right]\nonumber\\
&=\int_{\partial\mathcal{M}_5}d^4x\sqrt{-\gamma^{(0)}g^{uu}} \frac{g'_{xx}}{2g_{xx}}\phi^2 \biggr|_\epsilon^{u_H},
\label{2.14}
\end{align}
where we used relation (\ref{2.12}). Substituting (\ref{2.14}) into (\ref{2.13}), we obtain, up to $\mathcal{O}(\phi^2)$
\begin{align}
S_{\textrm{EH}}&\approx -\frac{1}{2\pi L^2G_5}V_4\int_\epsilon^{u_H}du\sqrt{-g^{(0)}}-\frac{1}{32\pi G_5} \int_{\mathcal{M}_5}d^5x\sqrt{-g^{(0)}}g_{(0)}^{MN}\partial_M\phi\partial_N\phi+\nonumber\\
&+\frac{1}{8\pi G_5}\int_{\partial\mathcal{M}_5}d^4x\sqrt{-\gamma^{(0)}g^{uu}}\left[\phi\phi ' + \frac{g'_{xx}}{4g_{xx}}\phi^2\right] \biggr|_\epsilon^{u_H}.
\label{2.15}
\end{align}
From (\ref{2.15}) we see that, as stated before, the bulk term of the disturbed EH action up to second order in the metric perturbation $\phi$ corresponds to the action for a massless scalar field in the undisturbed background $g_{MN}^{(0)}$, as is well-known. Hence, the linearized equation of motion for the mixed metric perturbation $\phi$ is just the massless Klein-Gordon equation in a curved background
\begin{align}
\nabla_{(0)}^2\phi=\frac{1}{\sqrt{-g^{(0)}}}\partial_M\left(\sqrt{-g^{(0)}}g_{(0)}^{MN}\partial_N\phi\right)=0\,.
\label{2.16}
\end{align}
Defining the Fourier transform\footnote{For the sake of notation simplicity, when necessary, we distinguish the position-space scalar field, $\phi(u,t,z)$, from its Fourier transform, $\phi(u,\omega,q)$, only by their arguments. Notice also that, since $\phi(u,t,z)$ is real-valued, we obtain the following reality condition in momentum space: $\phi(u,-\omega,-q)=\phi^*(u,\omega,q)$.} as 
\begin{align}
\phi(u,t,z)=\int_{-\infty}^\infty \frac{d\omega dq}{(2\pi)^2}e^{-i\omega t+iqz}\phi(u,\omega,q),
\label{2.19}
\end{align}
one finds in momentum space
\begin{align}
\partial_u\left(\sqrt{-g^{(0)}}g^{uu}\phi '(u,\omega,q)\right)=\sqrt{-g^{(0)}}\left(-g^{tt}\omega^2 +g^{xx}q^2\right)\phi(u,\omega,q)\,.
\label{2.17}
\end{align}
Integrating by parts the bulk piece of (\ref{2.15}) and making use of (\ref{2.16}), we put the EH action on-shell up to $\mathcal{O}(\phi^2)$
\begin{align}
S_{\textrm{EH,bdy}}^{\textrm{on-shell}}&\approx \frac{1}{2\pi L^2G_5}V_4 \lim_{u\rightarrow\epsilon}\int du\sqrt{-g^{(0)}} -\frac{1}{32\pi G_5}\int_{\partial\mathcal{M}_5}d^4x \lim_{u\rightarrow\epsilon} \sqrt{-\gamma^{(0)}g^{uu}}\left[3\phi\phi ' + \frac{g'_{xx}}{g_{xx}}\phi^2\right]^{(\textrm{on-shell})},
\label{2.18}
\end{align}
where, by following the prescription proposed in \cite{ss} for calculating the retarded propagator of the metric perturbation, we discarded the horizon contribution coming from the radial integration and took into account only the boundary contribution for the on-shell EH action.

Now we add (\ref{2.18}) to the contributions coming from the disturbed GHY and counterterm actions evaluated up to $\mathcal{O}(\phi^2)$ to find the following expression for the total regularized on-shell boundary action in momentum space\footnote{This is a rather lengthy calculation but it can be straightforwardly done with the help of a symbolic mathematical software such as Wolfram's Mathematica \cite{mathematica}. We used the EDCRGTC code \cite{code} written for Mathematica in order to deal with the lengthy tensor manipulations involved in the course of these calculations.} up to $\mathcal{O}(\phi^2,\omega^2,q^2)$
\begin{align}
S_{\textrm{reg}}(\epsilon)&\approx\frac{1}{16\pi G_5}\lim_{u\rightarrow\epsilon}\left\{\frac{V_4}{L^2}\left[-6L\sqrt{-\gamma^{(0)}} +8\int du\sqrt{-g^{(0)}}-\frac{L^2g_{xx}(g_{xx}g'_{tt}+3g_{tt}g'_{xx})}{\sqrt{g_{uu}g_{tt}g_{xx}}}\right]+\right.\nonumber\\
&+\int_{-\infty}^\infty \frac{d\omega dq}{(2\pi)^2}\frac{\phi(u,-k)}{2L^2}\left[\phi(u,k)\left(6L\sqrt{-\gamma^{(0)}}+ \frac{L^2g_{xx}(g_{xx}g'_{tt}+2g_{tt}g'_{xx})}{\sqrt{g_{uu}g_{tt}g_{xx}}}+\right.\right.\nonumber\\
&\left.\left.\left.+\frac{L^3q^2\sqrt{g_{tt}g_{xx}}}{2}-\frac{L^3\omega^2}{2} \sqrt{\frac{g_{xx}^3}{g_{tt}}}\right)+L^2\sqrt{-\gamma^{(0)}g^{uu}}\phi '(u,k)\right]\right\}.
\label{2.20}
\end{align}
One can formally solve (\ref{2.17}) in terms of the components of the undisturbed background metric, $g_{MN}^{(0)}(u)$, and the boundary value of the metric perturbation, $\varphi(\omega,q)$, by employing a perturbative expansion in $\omega\ll T$ and $q\ll T$. Such a solution up to $\mathcal{O}(\omega^2,q^2)$ is derived in details in Appendix \ref{apa} and the results near the boundary read
\begin{align}
\phi(\epsilon,k)&\approx\varphi(k),\label{2.21}\\
\phi '(\epsilon,k)&\approx\varphi(k)\left[\frac{i\omega}{4\pi T(u_H-\epsilon)}+f'(\epsilon,k)\right]\nonumber\\
&\approx \frac{\varphi(k)g_{xx}^{3/2}(u_H)}{\sqrt{-g^{(0)}(\epsilon)}g^{uu}(\epsilon)} \left[ i\omega + \omega^2 \int_{u_H}^\epsilon du\left(\frac{g_{xx}^{3/2}(u_H)}{\sqrt{-g^{(0)}}g^{uu}} -\frac{\sqrt{-g^{(0)}}g^{tt}}{g_{xx}^{3/2}(u_H)} \right)+q^2\int_{u_H}^\epsilon du\frac{\sqrt{-g^{(0)}}g^{xx}}{g_{xx}^{3/2}(u_H)} \right].
\label{2.22}
\end{align}
Substituting (\ref{2.21}) and (\ref{2.22}) into (\ref{2.20}) one obtains an expression for the total regularized on-shell boundary action up to $\mathcal{O}(\phi^2,\omega^2,q^2)$ written solely in terms of the components of the undisturbed background metric and the boundary value of the metric perturbation $\varphi(k)$.

The renormalized on-shell boundary gravity action is defined by
\begin{align}
S_{\textrm{ren}}\equiv\lim_{\epsilon\rightarrow 0}S_{\textrm{reg}}(\epsilon).
\label{2.23}
\end{align}

For completeness, let us also review some useful formulas for calculating the pressure, the entropy density, and the shear viscosity of the boundary plasma using the gauge/gravity duality. From (\ref{2.1}), we see that once we have extracted the retarded propagator from the renormalized on-shell boundary action by following the prescription proposed in \cite{ss}
\begin{align}
S_{\textrm{ren}}=\lim_{\epsilon\rightarrow 0}S_{\textrm{reg}}(\epsilon)&=-\frac{1}{2}\lim_{\epsilon\rightarrow 0}\int_{-\infty}^\infty \frac{d\omega dq}{(2\pi)^2}\varphi(-\omega,-q) \mathcal{F}(\omega,q;\epsilon) \varphi(\omega,q)+(\varphi\textrm{-independent terms});\nonumber\\
G_R^{xy,xy}(\omega,q)&\equiv\lim_{\epsilon\rightarrow 0}\mathcal{F}(\omega,q;\epsilon)
\label{2.24}
\end{align}
one can obtain the pressure as follows
\begin{align}
P=\lim_{k\rightarrow 0}G_R^{xy,xy}(k).
\label{2.25}
\end{align}
Since the perturbation-independent part of the on-shell boundary action gives $-PV_4$ \cite{brsss}, we can also calculate the pressure by using the following alternative formula
\begin{align}
P=-\lim_{\varphi\rightarrow 0}\frac{S_{\textrm{ren}}}{V_4}.
\label{2.26}
\end{align}
One can evaluate the entropy of the plasma by using the Bekenstein-Hawking's relation \cite{bek-haw-1,bek-haw-2}
\begin{align}
S=\frac{A_H}{4G_5},
\label{2.27}
\end{align}
where the ``area" of the horizon is given by
\begin{align}
A_H=\int_{\textrm{horizon}}d^3x \sqrt{g}\biggr|_{u=u_H\, \textrm{and}\, t \,\textrm{fixed}}=\sqrt{g_{xx}^3(u_H)}V_3,
\label{2.28}
\end{align}
and the entropy density reads
\begin{align}
s=\frac{S}{V_3}=\frac{\sqrt{g_{xx}^3(u_H)}}{4G_5}.
\label{2.29}
\end{align}

From (\ref{2.1}), one also obtains the following Kubo's formula for the shear viscosity
\begin{align}
\eta=-\lim_{q\rightarrow 0}\lim_{\omega\rightarrow 0}\textrm{Im}\left[\frac{\partial G_R^{xy,xy}(\omega,q)}{\partial\omega}\right].
\label{2.30}
\end{align}

\subsection*{Application: Thermal $\mathcal{N}=4$ Supersymmetric Yang-Mills (SYM)}
\label{sec2.1}

In this subsection we review the calculations for SYM at finite temperature \cite{brsss} whose gravity dual is defined over an AdS$_5$-Schwarzschild background with metric components 
\begin{align}
g_{uu}(u)=\frac{L^2}{u^2h(u)},\,\,\,g_{tt}(u)=\frac{L^2h(u)}{u^2},\,\,\,g_{xx}(u)=\frac{L^2}{u^2};\,\,\,h(u)=1 -\frac{u^4}{u_H^4}.
\label{2.31}
\end{align}
The relation between $G_5$ and the number of colors of the strongly coupled SYM plasma reads \cite{peet} 
\begin{align}
G_5=\frac{G_{10}}{\pi^3 L^5}=\frac{\pi L^3}{2N_c^2}\,.
\label{2.32}
\end{align}
By substituting (\ref{2.31}) and (\ref{2.32}) into (\ref{a5}) and (\ref{2.29}), one obtains, respectively
\begin{align}
T=\frac{1}{\pi u_H},\,\,\,s=\frac{N_c^2}{2\pi L^3}\frac{L^3}{{u_H}^3}=\frac{\pi^2 T^3 N_c^2}{2}.
\label{2.33}
\end{align}
Now we use (\ref{2.31}), (\ref{2.32}), and (\ref{2.33}) to calculate (\ref{2.22}) and, by substituting the result into (\ref{2.20}) and then evaluating (\ref{2.23}), we obtain the renormalized on-shell boundary action up to $\mathcal{O}(\phi^2,\omega^2,q^2)$ for SYM
\begin{align}
S_{\textrm{ren}}\approx -\frac{\pi^2T^4N_c^2}{8} V_4+\int_{-\infty}^\infty \frac{d\omega dq}{(2\pi)^2} \varphi(-k)\varphi(k) \frac{T^2N_c^2}{32}\left [q^2-2\pi^2T^2+2i\omega\pi T-\omega^2 (1-\ln 2)\right]\,.
\label{2.34}
\end{align}
Substituting (\ref{2.34}) into (\ref{2.24}), we obtain the graviton propagator up to $\mathcal{O}(\phi^2,\omega^2,q^2)$
\begin{align}
G_R^{xy,xy}(\omega,q)\approx -\frac{T^2N_c^2}{16}\left[q^2-2\pi^2T^2+2i\omega\pi T-\omega^2 (1-\ln 2)\right],
\label{2.35}
\end{align}
and by using (\ref{2.25}) (or also (\ref{2.26})), we obtain the pressure of the conformal strongly-coupled SYM plasma
\begin{align}
P=\frac{\pi^2T^4N_c^2}{8}.
\label{2.36}
\end{align}
The first order hydrodynamic transport coefficient $\eta$ for the SYM plasma is found by substituting (\ref{2.35}) into (\ref{2.30})
\begin{align}
\eta=\frac{\pi T^3N_c^2}{8}.
\label{2.37}
\end{align}
From (\ref{2.33}) and (\ref{2.37}), one obtains the famous ratio
\begin{align}
\frac{\eta}{s}=\frac{1}{4\pi},
\label{2.38}
\end{align}
which is actually valid for a broad class of gravity duals \cite{kss,Buchel:2003tz}. The second order hydrodynamic transport coefficients $\kappa$ and $\tau_\pi$ for the SYM plasma are found by substituting (\ref{2.35}) into (\ref{2.3}) and (\ref{2.4}), respectively
\begin{align}
\kappa&=\frac{T^2N_c^2}{8},\label{2.39}\\
\tau_\pi&=\frac{2-\ln 2}{2\pi T}.\label{2.40}
\end{align}
These results for the conformal SYM plasma were originally obtained in \cite{brsss}.

\subsection{General holographic formulas for $\kappa$ and $\tau_\pi$ in Einstein+Scalar gravity duals}
\label{sec3}

In this section we derive general formulas for $\kappa$ and $\tau_\pi$ which are valid also for systems where the metric couples to matter fields in the bulk (we shall focus here on the case where there is a scalar field in the bulk).

For geometries corresponding to solutions of the field equations which take into account the backreaction of matter fields in the bulk, the pure gravity regularized action (\ref{2.20}) shall feature, in general, temperature-independent divergences as one takes the limit $\epsilon\rightarrow 0$. For instance, this happens in Einstein+Scalar models \cite{ihqcd-1,ihqcd-2,hot-ihqcd,gubser-nellore,Li:2011hp}. For this kind of system, the general procedure of holographic renormalization is discussed in \cite{ren5}. Here, since we are only interested in evaluating hydrodynamic transport coefficients\footnote{Notice that the Kubo's formulas for transport coefficients are defined in terms of momentum derivatives of retarded correlators and, therefore, momentum-independent terms in the on-shell action do not contribute in such calculations.}, instead of dealing with a more complicated regularized action, we are going to apply a physical prescription which will allow us to obtain quite general formulas for $\kappa$ and $\tau_\pi$ through a simple procedure. This prescription is based on three main facts:
\begin{enumerate}
\item The equation of motion for the $xy$-component of the metric perturbation depends only on the background metric \cite{pss,kss} and, therefore, the solution (\ref{2.22}) remains valid also for Einstein+Scalar actions. 

\item Ultraviolet divergences are temperature-independent and, consequently, one can remove these divergences by subtracting contributions evaluated at different temperatures. 

\item The coefficients $\kappa$ and $\tau_\pi \eta$, which appear in the gradient expansion, must vanish at sufficiently low temperatures (as one approaches the vacuum).

\end{enumerate}

Let us first discuss the coefficient $\kappa$ in \eqref{2.3}. We begin by defining the following regularized quantity (see eqs. \eqref{2.20} and \eqref{2.24})
\begin{align}
\kappa_\epsilon := -\lim_{q\rightarrow 0}\lim_{\omega\rightarrow 0}\frac{\partial^2}{\partial q^2}\mathcal{F}(\omega,q;\epsilon)
=\frac{1}{8\pi G_5}\left[\frac{L^3}{2\epsilon^2}+\int_{u_H}^\epsilon du\sqrt{-g^{(0)}}g^{xx}\right],
\label{3.1}
\end{align}
and then we take the following expression, which is free of ultraviolet divergences 
\begin{align}
\kappa(T)=& \lim_{\epsilon\rightarrow 0}\left(\kappa_\epsilon(T) -\kappa_\epsilon(T_{\textrm{high}})\right) +\kappa_{\textrm{SYM}}(T_{\textrm{high}})-\kappa_0& \nonumber\\
\approx &\frac{1}{8\pi G_5}\left[\int_{u_H(T)}^\epsilon du\,\sqrt{-g^{(0)}} g^{xx}\biggr|_{u_H(T)} -\int_{u_{\textrm{high}}}^\epsilon du\,\sqrt{-g^{(0)}} g^{xx}\biggr|_{u_{\textrm{high}}}\right]\biggr|_{\epsilon \ll 1}& \nonumber \\  &+ \kappa_{\textrm{SYM}}(T_{\textrm{high}})-\kappa_0\,,&
\label{3.2}
\end{align}
where $T_{\textrm{high}}$ is a temperature that is sufficiently large so that we are near the ultraviolet fixed point and the temperature-dependent part\footnote{Which is independent of the ultraviolet cutoff $\epsilon$ as mentioned before.} of $\kappa_\epsilon(T_{\textrm{high}})$ approaches $\kappa_{\textrm{SYM}}(T_{\textrm{high}})$, $u_{\textrm{high}}=u_H(T_{\textrm{high}})$, and $\kappa_0$ is a constant to be subtracted (generally by numerical inspection) in order to ensure that $\kappa(T_{\textrm{min}})=0$ where $T_{\textrm{min}}$ is the lowest temperature considered in our numerical calculations ($T_{\textrm{min}} \sim 10$ MeV). From (\ref{2.32}) and (\ref{2.39}), one finds
\begin{align}
\kappa_{\textrm{SYM}}=\frac{T^2N_c^2}{8}=\frac{\pi T^2 L^3}{16 G_5}\,.
\label{3.3}
\end{align}
Analogously, for $\tau_\pi$ in \eqref{2.4} we begin by defining
\begin{align}
\tau_\pi&=\frac{1}{2\eta} \left(\Omega-\kappa+T\frac{d\kappa}{dT}\right),\label{3.4}\\
\Omega_\epsilon&:=\lim_{q\rightarrow 0}\lim_{\omega\rightarrow 0}\frac{\partial^2}{\partial \omega^2}\mathcal{F}(\omega,q;\epsilon) = \nonumber\\
&=\frac{1}{8\pi G_5}\left[\frac{L^3}{2\epsilon^2}+g_{xx}^{3/2}(u_H)\int_\epsilon^{u_H} du \left(\frac{g_{xx}^{3/2}(u_H)}{\sqrt{-g^{(0)}}g^{uu}} -\frac{\sqrt{-g^{(0)}}g^{tt}}{g_{xx}^{3/2}(u_H)}\right) \right],
\label{3.5}
\end{align}
and then we evaluate the UV finite expression
\begin{align}
\Omega(T)&=\lim_{\epsilon\rightarrow 0}\left(\Omega_\epsilon(T) -\Omega_\epsilon(T_{\textrm{high}})\right) +\Omega_{\textrm{SYM}}(T_{\textrm{high}})-\Omega_0\nonumber\\
&\approx \frac{1}{8\pi G_5}\left(g_{xx}^{3/2}(u_H)\int^{u_H(T)}_\epsilon du\left[\frac{g_{xx}^{3/2}(u_H)}{\sqrt{-g^{(0)}}g^{uu}}-\frac{\sqrt{-g^{(0)}}g^{tt}}{g_{xx}^{3/2}(u_H)}\right]\biggr|_{u_H(T)} +\right.\nonumber\\
&\left. -g_{xx}^{3/2}(u_{\textrm{high}})\int^{u_{\textrm{high}}}_\epsilon du\left[\frac{g_{xx}^{3/2}(u_{\textrm{high}})}{\sqrt{-g^{(0)}}g^{uu}}- \frac{\sqrt{-g^{(0)}}g^{tt}}{g_{xx}^{3/2}(u_{\textrm{high}})} \right]\biggr|_{u_{\textrm{high}}}\right)\biggr|_{\epsilon \ll 1} +\Omega_{\textrm{SYM}}(T_{\textrm{high}})-\Omega_0,
\label{3.6}
\end{align}
where
\begin{align}
\Omega_{\textrm{SYM}}=\frac{T^2N_c^2}{8}[1-\ln 2]=\frac{\pi T^2 L^3}{16 G_5}[1-\ln 2]\,,
\label{3.7}
\end{align}
and $\Omega_0$ is a constant that is subtracted to ensure that $\tau_\pi(T_{\textrm{min}}) \eta(T_{\textrm{min}}) = 0$.

Eqs.\ (\ref{3.2}), (\ref{3.4}), and (\ref{3.6}) can be used to compute the transport coefficients $\kappa$ and $\tau_\pi$ for Einstein+Scalar actions. In the next section we employ them to numerically evaluate $\kappa$ and $\tau_\pi$ in a holographic model for $(2+1)$-flavor QCD at finite temperature and zero baryon density proposed by Gubser and Nellore \cite{gubser-nellore}.

\section{Thermodynamics of the Einstein+Scalar holographic model}
\label{sec4}

In this section we shall briefly review a bottom-up holographic model \cite{gubser-nellore} which is built to provide a quantitative description of the thermodynamical properties of $(2+1)$-flavor QCD as seen on the lattice\footnote{For holographic studies of the deconfinement phase transition see, for instance, \cite{Herzog:2006ra,BallonBayona:2007vp}.}. In the next section we will apply the results of the preceding sections to compute all of the first order transport coefficients and most of the second order ones, presenting estimates for these coefficients that may be relevant for the hydrodynamic simulations of the QGP. 

The Einstein+Scalar holographic model has been successfully used in several other works to understand the temperature dependence of quantities evaluated near the crossover phase transition of the QGP such as the bulk viscosity \cite{Gubser:2008yx}, the expectation value of the Polyakov loop \cite{Noronha:2009ud,Noronha:2010hb}, the energy loss of heavy and light quarks \cite{Ficnar:2010rn,Ficnar:2011yj,Ficnar:2012yu}, the electric conductivity \cite{Finazzo:2013efa}, and the Debye screening mass \cite{Finazzo:topub}.

The bulk action for the Einstein+Scalar model considered in \cite{gubser-nellore} is given by
\begin{equation}
S_{\textrm{ES}}^{(\textrm{bulk})}=\frac{1}{16 \pi G_5}\int_{\mathcal{M}_5} d^5x \sqrt{-g}\left[R-\frac{(\partial_M \Phi)^2}{2}-V(\Phi)\right],
\label{4.1}
\end{equation}
where $\Phi$ is the scalar field and $V(\Phi)$ is the scalar potential, which we shall specify below. We note that this model cannot be used to describe aspects of the QCD phase transition that are directly related to chiral symmetry\footnote{For holographic bottom-up models that deal with effects of chiral symmetry see, for instance, \cite{Li:2013oda,Alho:2013hsa,Bartz:2014oba}.}. For the purpose of solving the equations of motion, we use the following Ansatz for the metric (in the so-called Gubser gauge) \cite{gubser-nellore} (see also \cite{Finazzo:topub} for more details)
\begin{equation}
\label{eq:gubsergauge}
ds^2=e^{2A(\Phi)}\left(-h(\Phi)dt^2+dx_i^2\right)+e^{2B(\Phi)}\frac{d\Phi^2}{h(\Phi)},
\end{equation}
where the scalar field itself is considered as the radial coordinate. In order to have a black brane background, we require that $h(\Phi)$ has a simple zero at $\Phi=\Phi_H$. We also impose that the metric is asymptotically $\mathrm{AdS}_5$ and that $\Phi$ is associated with a relevant operator in the gauge theory \cite{Gubser:2008yx}, which implies that $\Phi \to 0$ at the boundary. The entropy density is given by the Bekenstein-Hawking formula (\ref{2.29})
\begin{equation}
s=\frac{e^{3A(\Phi_H)}}{4G_5},
\label{4.3}
\end{equation}
and the corresponding temperature of the black brane is given by (\ref{a5})
\begin{equation}
T=e^{A(\Phi_H)-B(\Phi_H)} \frac{| h'(\Phi_H) |}{4\pi}\,.
\label{4.4}
\end{equation}

\begin{figure}
\centering
  \includegraphics[width=.5\linewidth]{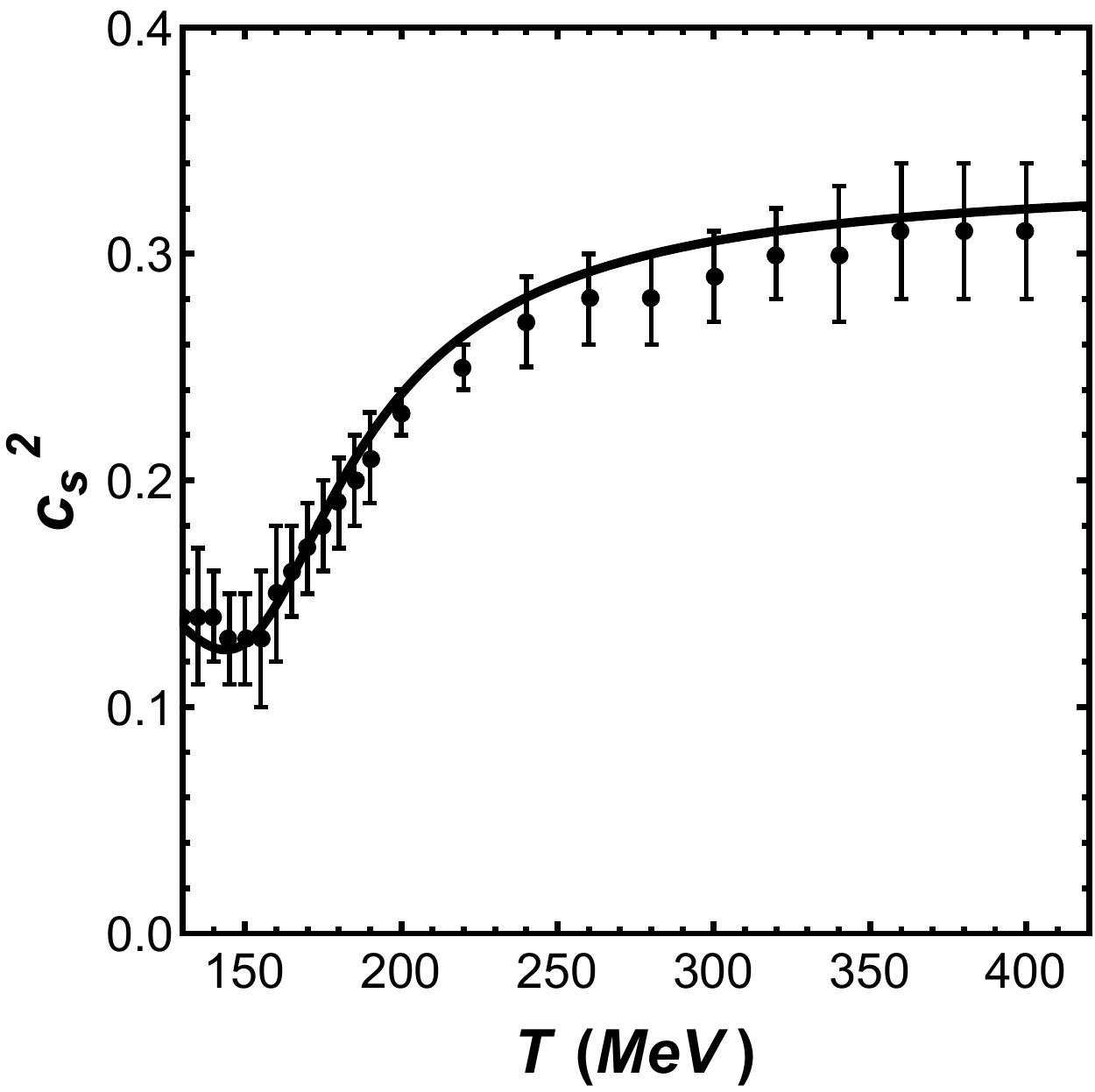}
  \caption{$c_s^2$ as a function of the temperature $T$ for the bottom-up holographic model (solid curve) and the corresponding lattice results for $(2+1)$-flavor QCD from \cite{fodor2012}.}
  \label{fig:cs2}      
\end{figure}

\begin{figure}
\centering
  \includegraphics[width=.5\linewidth]{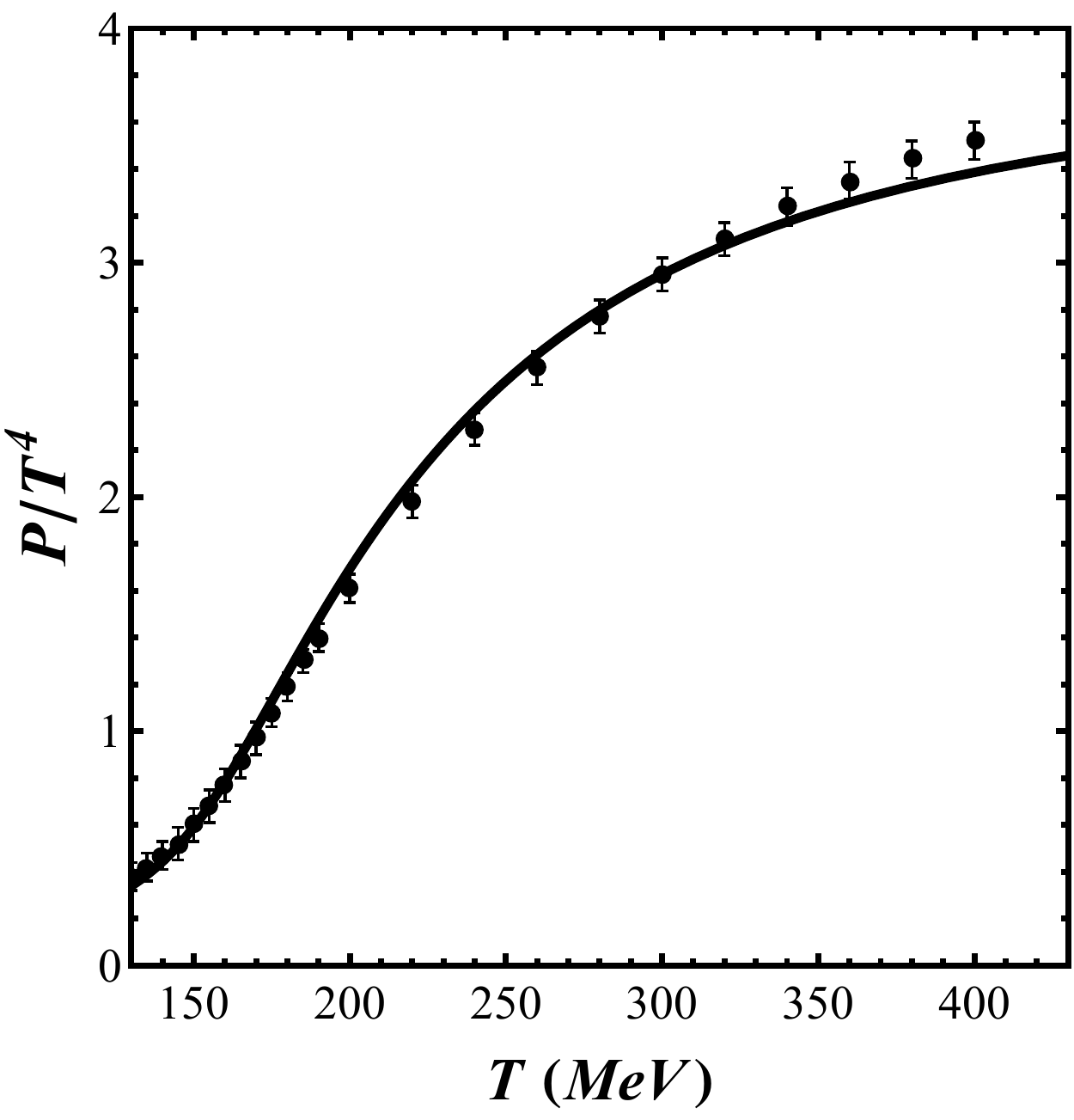}
  \caption{$P/T^4$ as a function of the temperature $T$ for the bottom-up holographic model (solid curve) and the corresponding lattice results for $(2+1)$-flavor QCD from \cite{fodor2012}.}
  \label{fig:pressure}      
\end{figure}

The scalar potential $V(\Phi)$ is chosen in order to mimic the thermodynamics of $(2+1)$-flavor QCD at zero baryon chemical potential as calculated on the lattice \cite{fodor2012}, represented by $c_s^2(T)=\frac{d\ln T}{d\ln s}$, the square of the speed of sound in the plasma as a function of the temperature. The potential we use is
\begin{equation}
V(\Phi)=\frac{-12 \cosh \gamma \Phi + b_2 \Phi^2 + b_4 \Phi^4 + b_6 \Phi^6}{L^2},
\label{4.5}
\end{equation} 
with $\gamma=0.606$, $b_2=0.703$, $b_4=-0.1$, $b_6=0.0034$ (and the asymptotic $\mathrm{AdS}_5$ radius fixed as $L=1$). The temperature scale is chosen in order to match the minimum of the speed of sound $c_s^2$ computed holographically with that found on the lattice (we take this minimum to be at $143.8 \, \mathrm{MeV}$). The thermodynamics of $(2+1)$-flavor QCD is also used to fix $G_5$ by matching lattice data for $P/T^4$; this gives $G_5 = 0.5013$. In Figs.\ \ref{fig:cs2} and \ref{fig:pressure} we compare the lattice results with the results of this holographic model for $c_s^2$ and $P/T^4$ as a function of the temperature $T$. 

The model is able to describe the temperature region near the minimum of the speed of sound but the agreement does not persist at very high temperatures, which is expected since the model remains strongly interacting in this case while QCD is asymptotically free. Moreover, even though this model does not have the correct (hadronic) degrees of freedom at low temperatures, the temperature dependence of the thermodynamical quantities do follow lattice data even for $T \sim 130$ MeV. The model may then be useful precisely in the temperature region $T \sim 130-450$ MeV where a purely hadronic description is not adequate and the temperature may not be high enough to warrant a simple description using perturbative QCD\footnote{We note, however, that {\it non-perturbative} weak coupling approaches such as the one pursued in \cite{Mogliacci:2013mca,Haque:2014rua} do a fairly good job at describing the thermodynamic properties of QCD found on the lattice at high temperatures. The motivation for finding a holographic description of the strongly-coupled QGP relies on the fact that holography is not only able to describe the thermodynamics near the phase transition but it also allows for the direct calculation of non-equilibrium properties, such as transport coefficients, within the same setup.}.

\section{Holographic calculation of the transport coefficients}\label{transportsec}
\label{sec4.2}

\subsubsection*{Coefficient $\kappa$}
\label{sec4.2.1}

The transport coefficients that we will compute in this section appear in the second-order gradient expansion equations in (\ref{defineshearstress}) and (\ref{definebulk}). We first compute the coefficient $\kappa$ using \eqref{3.2}. The first step is to fix a $u_{\mathrm{high}}$ at which the geometry computed by solving the equations of motion has already reached its $AdS_5$ asymptotics. For numerical integration, it is necessary to keep the cutoff $\epsilon$ in \eqref{3.2}. A value for the cutoff that is too small leads to truncation errors due to the subtraction of two small numbers in \eqref{3.2}, which gives an artificial numerical divergence. Thus, to rule out errors introduced due to truncation, one should compute the integral in \eqref{3.2} with a range of choices for $\epsilon$ and search for a reasonable range that does not introduce numerical divergences in the integral (which gives a lower bound for $\epsilon$) and also satisfies $\epsilon < u_{\mathrm{high}}$ (which gives an upper bound for $\epsilon$). We found that the optimal region for numerical calculations is $10^{-3} < \epsilon < 10^{-1}$ and we have chosen in this work $\epsilon = 10^{-2}$ for the calculation of $\kappa$. We have chosen $u_{\mathrm{high}} = 0.201$, which corresponds to $T_{\mathrm{high}} = 7.813 \,T_c$ (where $T_c = 143.8 \, \mathrm{MeV}$). We have checked that at this high temperature both the thermodynamics as well as the transport coefficients have matched their conformal plasma limits.

\begin{figure}
\centering
  \includegraphics[width=.5\linewidth]{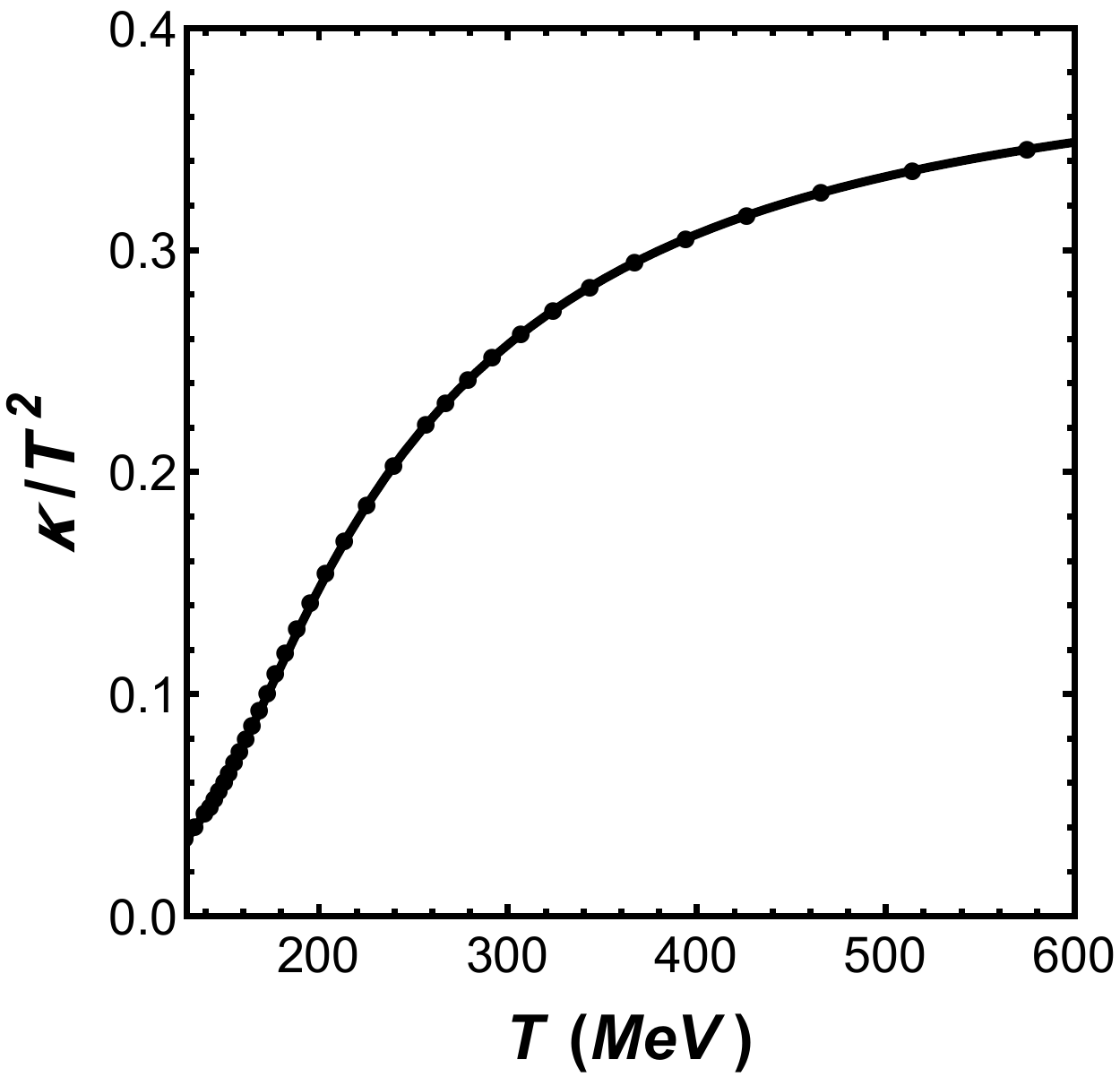}
  \caption{$\kappa/T^2$ as a function of the temperature $T$ for the bottom-up holographic model. The black points correspond to the numerical results from the model while the solid black line corresponds to the fit in Eq.\ (\ref{eq:kappafit}) with the parameters in Table \ref{tab:parameterskappa}.}
  \label{fig:kappa}      
\end{figure}

Proceeding as discussed in the previous paragraph, we determine $\kappa$ as a function of $u_H(T)$ for the chosen $u_{\mathrm{high}}$. Using the conformal result at strongly coupling \eqref{3.3} and the value of $G_5$ determined in the previous section, we can determine the dimensionless ratio $\kappa/T^2$. In Fig.\ \ref{fig:kappa} we show the numerical results for $\kappa/T^2$ as a function of $T$. Moreover, we see that $\kappa/T^2$ approaches the conformal limit from below rising monotonically with $T$. Our results are consistent with the lattice results in \cite{lattice-kappa} where the authors obtained $\kappa/T^2 \sim 0.36 (15)$ for $T = 2-10 \,T_c$ for a pure glue $SU(3)$ plasma (in this case $T_c$ is the critical temperature for the first-order deconfinement phase transition). However, in our model we are able to obtain an estimate for the behavior of $\kappa$ near the crossover transition of (2+1)-flavor QCD.

For further use, we also present a fit to our calculated points. We use as a fitting function the seven-parameter fit
\begin{equation}
\label{eq:kappafit}
\frac{\kappa}{T^2}\left(x=\frac{T}{T_c} \right) = \frac{a}{1+ e^{b\left(c-x\right)} + e^{d\left(e-x\right)} + e^{f\left(g-x\right)}},
\end{equation}
where $a$ to $g$ are dimensionless fit parameters and $T_c = 143.8$ MeV, as remarked before. A five parameter fit using parameters from $a$ to $e$ already yields good results - the two extra parameters $f$, $g$ are used to provide a closer match to the points. This function corresponds to a modified Fermi-Dirac distribution. In Table \ref {tab:parameterskappa} we present the parameters for the fit. These fit parameters provide a good description of our numerical data as shown in Fig.\ \ref{fig:kappa}. 

\begin{table}[htp]
\caption{Parameters for the fit of $\kappa/T^2$ using Eq.\ \eqref{eq:kappafit}.}
\begin{center}
\begin{tabular}{cccccccc}

\hline
\hline

$a$ & $b$ & $c$ & $d$ & $e$ & $f$  & $g$ \\ \hline
0.3817 & 2.047 & 1.274 & 6.0545 & 1.231 & 0.5438 & -0.2076 \\ \hline\hline

\end{tabular}
\end{center}
\label{tab:parameterskappa}
\end{table}

\subsubsection*{Coefficient $\tau_{\pi}$}
\label{sec4.2.2}

With the same remarks as in the preceding section one can evaluate $\tau_{\pi}$ using \eqref{3.4} and \eqref{3.6}. The procedure is similar to the evaluation of $\kappa$. The same $u_{\mathrm{high}}$ was chosen while in the present case $\epsilon = 2 \times 10^{-2}$ - the integrals in \eqref{3.6} are more complicated than the integrals in \eqref{3.2} and also more sensitive to the choice of the cutoff.

\begin{figure}
\centering
  \includegraphics[width=.5\linewidth]{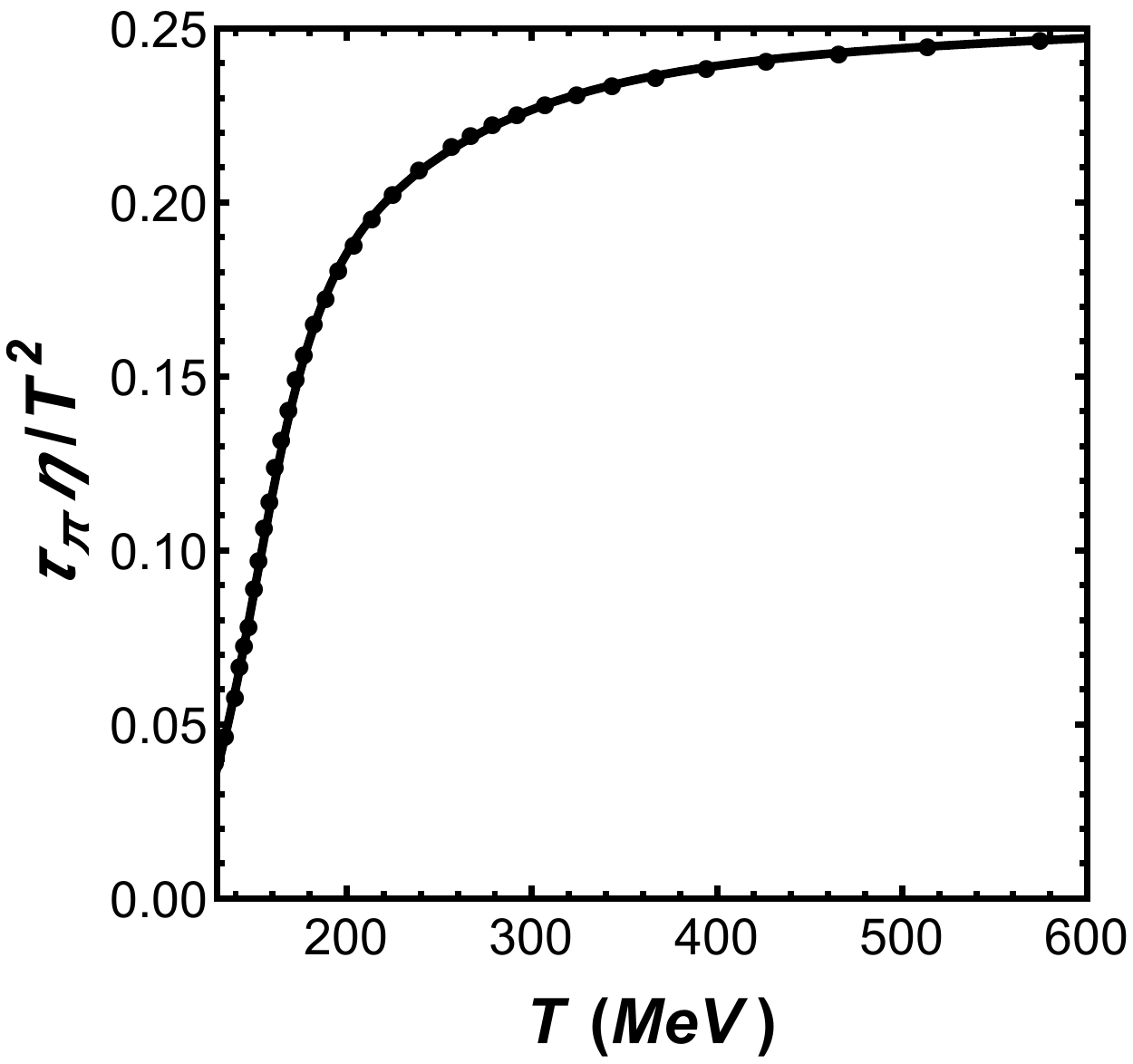}
  \caption{$\tau_{\pi} \eta /T^2$ as a function of the temperature $T$ for the bottom-up holographic model. The black points correspond to the numerical results from the model while the solid black line corresponds to the fit in Eq.\ (\ref{eq:kappafit}) with the parameters from Table \ref{tab:parameterstaupi}.}
  \label{fig:taupieta}      
\end{figure}

In Fig.\ \ref{fig:taupieta} we show the numerical results for $\tau_{\pi} \eta/T^2$ (this is a convenient choice since, by combining the conformal results \eqref{2.37} and \eqref{2.40}, we see that $\tau_{\pi} \eta \sim T^2$ for a conformal plasma) as a function of $T$. We see the same general behavior as seen for $\kappa/T^2$ or $P/T^4$, with a marked increase near the crossover region. The transport quantity $\tau_{\pi} \eta/T^2$ approaches its conformal limit ($\sim 0.255$) from below.
We were able to fit these data points using the same parametrization as used for $\kappa/T^2$ in Eq.\ \eqref{eq:kappafit} - the fit parameters are shown in Table \ref{tab:parameterstaupi}. 
\begin{table}[htp]
\caption{Parameters for the fit of $\tau_{\pi} \eta/T^2$ using Eq.\ \eqref{eq:kappafit}. }
\begin{center}
\begin{tabular}{cccccccc}

\hline
\hline

$a$ & $b$ & $c$ & $d$ & $e$ & $f$  & $g$ \\ \hline
0.2664 & 2.029 & 0.7413 & 0.1717 & -10.76 & 9.763 & 1.074 \\ \hline\hline

\end{tabular}
\end{center}
\label{tab:parameterstaupi}
\end{table}

\begin{figure}
\centering
  \includegraphics[width=.5\linewidth]{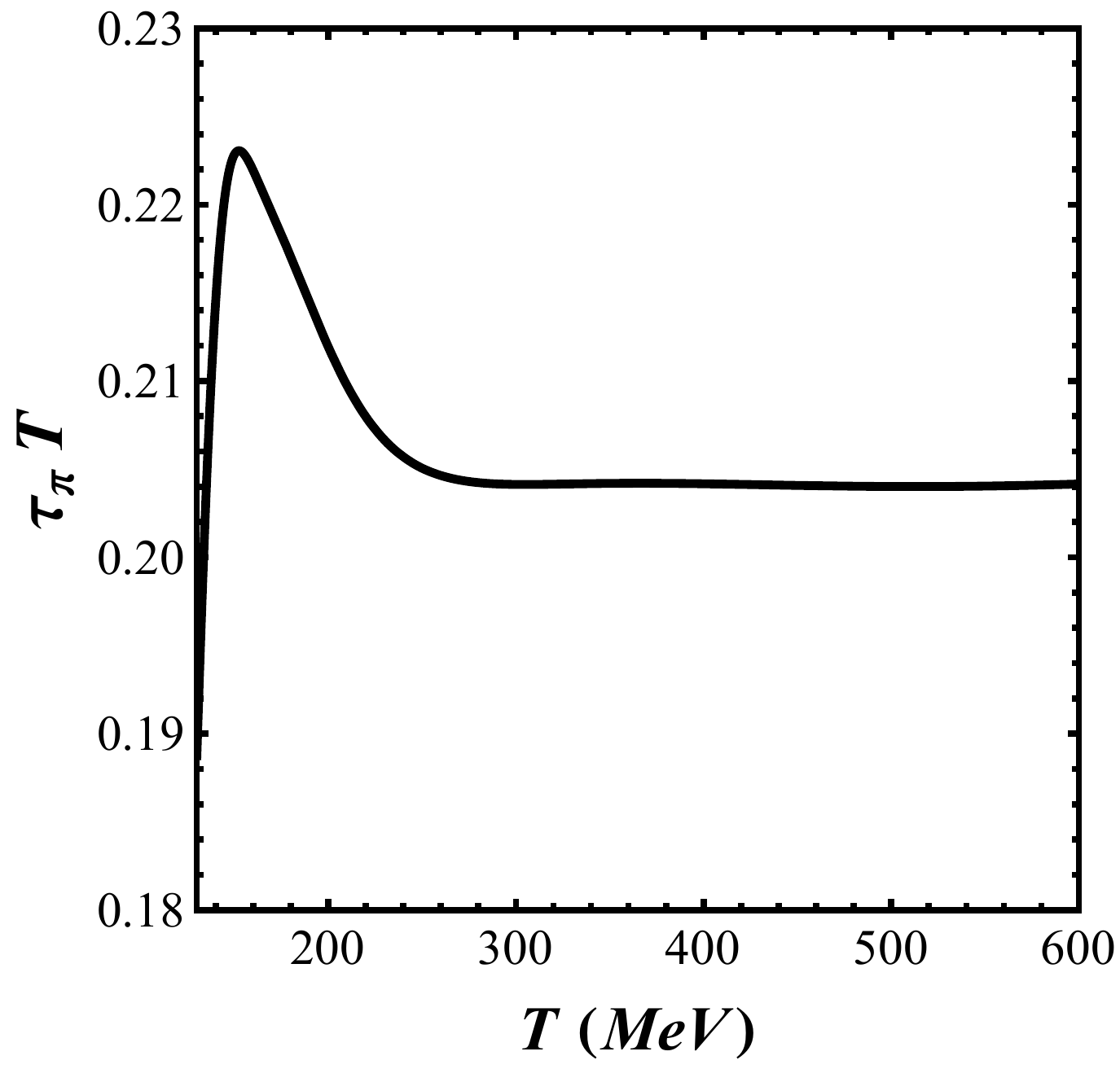}
  \caption{$\tau_{\pi} T$ as a function of the temperature $T$ for the bottom-up holographic model.}
  \label{fig:taupiT}      
\end{figure}

For convenience, we show in Fig.\ \ref{fig:taupiT} the quantity $\tau_\pi T$. One can see that this quantity approaches its conformal value, $(2-\ln 2)/(2\pi)$, for $T > 300$ MeV while it displays a peak near the transition.

%

\subsubsection*{Coefficient $\kappa^*$}

\begin{figure}
\centering
  \includegraphics[width=.5\linewidth]{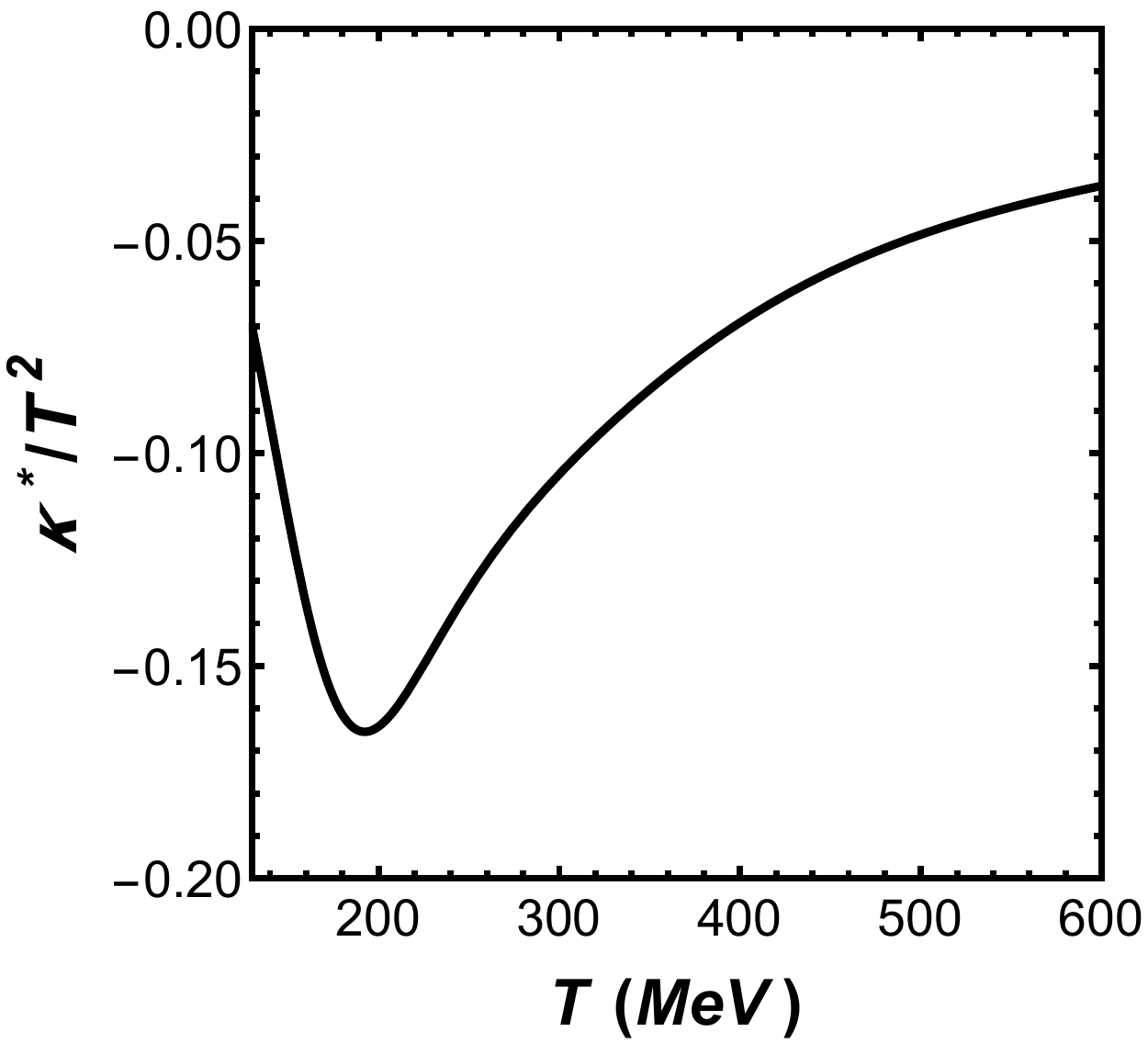}
  \caption{$\kappa^*/T^2$ as a function of the temperature $T$ for the bottom-up holographic model.}
  \label{fig:kappastar} 
\end{figure}

We can now use our results for $\kappa$ to evaluate several other transport coefficients of second order non-conformal hydrodynamics. Since these coefficients involve derivatives of $\kappa$ with respect to the temperature, up to third order (in the case of the coefficient $\xi_4$ in \eqref{eq:xi42}, see below), to avoid discretization errors in the computation of the numerical derivatives we will use the parametrization given by Eq.\ \eqref{eq:kappafit} with the parameters displayed in Table \ref{tab:parameterskappa}. The numerical results for the coefficient $\kappa^*$, defined by (\ref{2.2}) and computed this way, are shown in Fig.\ \ref{fig:kappastar}. One can see that $\kappa^*/T^2 \to 0$ as $T \to \infty$ - this can also be checked directly from the expression used for the fit, Eq. \eqref{eq:kappafit}. We note that this is in agreement with the fact that this coefficient vanishes for a conformal plasma. Also, this coefficient has a very sharp dependence with the temperature near the phase transition (following the behavior displayed by $c_s^2$) and $\kappa^* < 0$ for all the temperatures used here.

\subsubsection*{Coefficient $\xi_5$}

\begin{figure}
\centering
  \includegraphics[width=.5\linewidth]{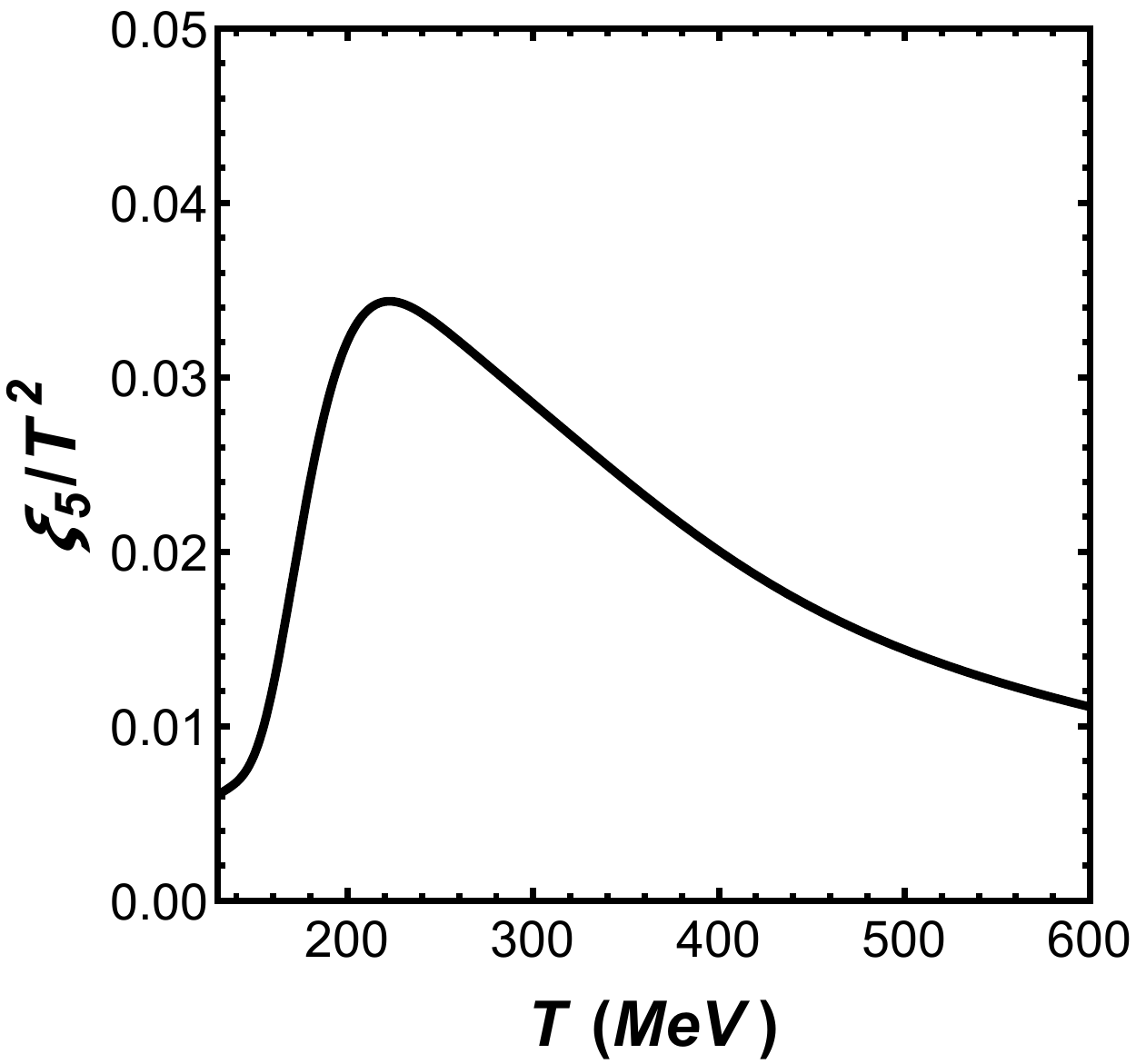}
  \caption{$\xi_5/T^2$ as a function of the temperature $T$ for the bottom-up holographic model.}
  \label{fig:xi5}      
\end{figure}

Another second order transport coefficient that we can directly evaluate is $\xi_5$, which is determined by the following constraint equation \cite{hydro1,hydro2,hydro3,moore}
\begin{equation}
\xi_5 = \frac{1}{2} \left(c_s^2 T \frac{d \kappa}{dT} - c_s^2 \kappa - \frac{\kappa}{3} \right)\,.
\label{4.6}
\end{equation}
In conformal hydrodynamics one finds that $\xi_5 = 0$. We evaluated $\xi_5/T^2$ as a function of the temperature using the equation above and the result can be found in Fig.\ \ref{fig:xi5}. One can see that $\xi_5/T^2$ has a broad peak in the phase transition around $T \sim 150-250 \, \mathrm{MeV}$ and decreases at high temperatures.

\subsubsection*{Shear and bulk viscosities}
\label{sec4.3}

For any isotropic holographic model with an effective gravitational action with at most two derivatives, the shear viscosity satisfies the ratio $\eta/s=1/(4\pi)$ \cite{kss} and this is the case of our model\footnote{We note, however, that the shear viscosity does not obey this ratio if higher order derivative corrections are included in the action \cite{Buchel:2004di,Kats:2007mq,Brigante:2007nu} or if the plasma is not isotropic \cite{Natsuume:2010ky,Erdmenger:2010xm,Rebhan:2011vd,Mamo:2012sy,Critelli:2014kra}.}.

We can now compute the bulk viscosity $\zeta$ using the results of \cite{prescription-justification-2,Gubser:2008yx}. This transport coefficient has attracted some attention recently due to its interplay with shear viscosity effects in event-by-event hydrodynamic simulations \cite{Noronha-Hostler:2013gga,Noronha-Hostler:2014dqa,Rose:2014fba,Gardim:2014tya}. The bulk viscosity is given by the Kubo formula
\begin{equation}
\label{eq:kubobulk}
\zeta = -\frac{4}{9} \lim_{\omega \to 0} \frac{1}{\omega}\,\mathrm{Im}\left[G_R (\omega, \vec{q} = \vec{0})\right],
\end{equation}
which is defined in terms of the retarded propagator of the spatial trace of the boundary stress-energy tensor
\begin{equation}
\label{eq:retzeta}
G_R (\omega, \vec{q}) \equiv -i \int_{\mathbb{R}^{1,3}} d^4x\, e^{i (\omega t - \vec{q}\cdot \vec{x})} \theta(t) \left\langle \left[ \frac{1}{2}T_a^a(t,\vec{x}), \frac{1}{2}T_b^b(0,\vec{0}) \right] \right\rangle\,.
\end{equation}
Holographically, this coefficient is computed considering fluctuations of the $xx$-component of the metric\footnote{In \cite{prescription-justification-2} the authors have shown that in the Gubser $\phi = r$ gauge the $h_{xx}$ fluctuation decouples from the other fluctuations, which means that we can examine this channel in separate in order to compute the bulk viscosity.}, $h_{xx}$. The equation of motion for the perturbation $\psi \equiv h^x_x = e^{-2A(\phi)} h_{xx}$ is given by
\begin{equation}
\label{eq:zetaedo}
\psi'' + \left(\frac{1}{3A'} + 4 A' - 3 B' + \frac{h'}{h} \right) \psi' + \left(\frac{e^{-2A+2B}}{h^2} \omega^2 - \frac{h'}{6hA'} + \frac{h'B'}{h} \right) \psi = 0,
\end{equation}
where the prime denotes a $\phi$-derivative.

As usual, in order to apply the real time prescription for the holographic computation of retarded correlators, we consider the infalling wave condition at the horizon $\phi = \phi_H$
\begin{equation}
\psi (\phi \to \phi_H) \approx C e^{i \omega t} |\phi - \phi_H|^{-\frac{i\omega}{4\pi T}},
\label{4.18}
\end{equation}
with the normalization condition $\psi (\phi \to 0) = 1$ at the boundary. The real time prescription implies that the imaginary part of the retarded correlator is given by
\begin{equation}
\label{eq:ImG}
\mathrm{Im}\left[G_R (\omega)\right] = - \frac{\mathcal{F}(\omega,\phi)}{16 \pi G_5},
\end{equation}
where $\mathcal{F}(\omega, \phi)$ is a conserved flux in the radial direction
\begin{equation}
\mathcal{F}(\omega, \phi) = \frac{e^{4A-B} h}{4 A'^2} |\mathrm{Im}\left[\psi^*\psi'\right]|,
\end{equation}
which can then be conveniently evaluated at the horizon\footnote{We expand $h(\phi)$ around $\phi=\phi_H$ to obtain $h(\phi\rightarrow\phi_H)\approx h'(\phi_H)(\phi-\phi_H)$. We also make use of (\ref{4.4}) to obtain $h'(\phi_H)=4\pi T/e^{A(\phi_H)-B(\phi_H)}$.}
\begin{align}
\mathcal{F}(\omega, \phi\rightarrow\phi_H) &\approx \frac{e^{3A(\phi_H)}}{4A'(\phi_H)^2}\, \frac{e^{A(\phi_H)-B(\phi_H)}}{4\pi T}\, h'(\phi_H)\, \omega |C|^2 \nonumber\\
&\approx \frac{\omega e^{3 A(\phi_H)} |C|^2}{4A'(\phi_H)^2}. \label{eq:boundF}
\end{align}

\begin{figure}
\centering
  \includegraphics[width=.5\linewidth]{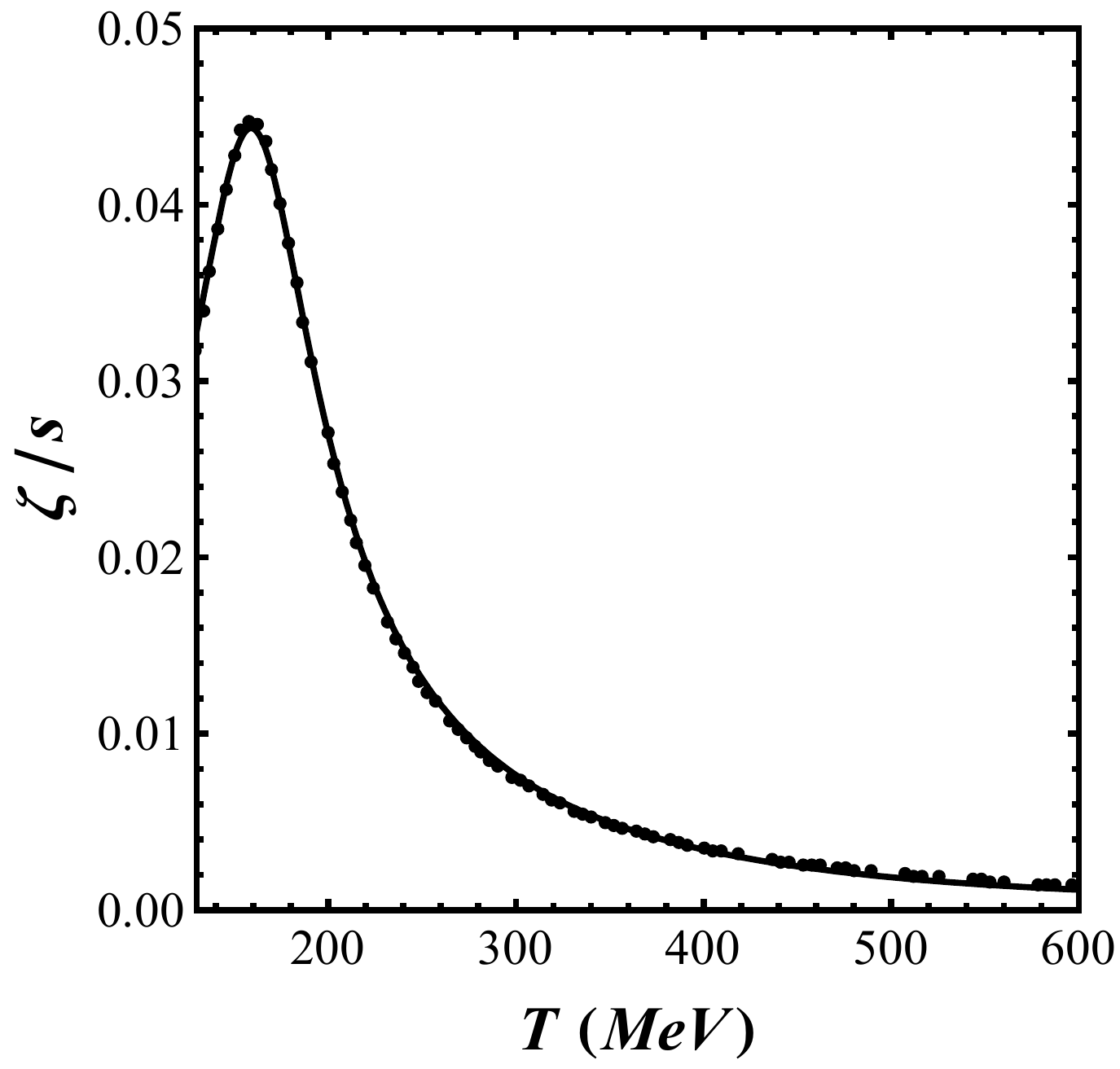}
  \caption{$\zeta/s$ as a function of the temperature $T$ for the bottom-up holographic model. The black points correspond to our numerical results while the black curve is the fit in Eq.\ \eqref{eq:zetafit} with the parameters in Table \ref{tab:parameterszeta}.}
  \label{fig:zeta}      
\end{figure}

Substituting Eq.\ \eqref{eq:boundF} into Eq.\ \eqref{eq:ImG} and then into Eq.\ \eqref{eq:kubobulk}, one finds
\begin{equation}
\frac{\zeta}{s} = \frac{\eta}{s}\,|C|^2 \frac{V'(\phi_H)^2}{V(\phi_H)^2}\,,
\label{zetaovereta}
\end{equation}
where we used Eqs.\ \eqref{4.3}, \eqref{2.38}, and also the relation $A'(\phi_H) = -V(\phi_H)/3V'(\phi_H)$, which can be derived using Einstein's equations for the background metric. Therefore, in order to compute $\zeta/s$  from (\ref{zetaovereta}), one only needs to evaluate (\ref{4.18}) in the limit of zero frequency $C = \lim_{\omega\rightarrow 0} \psi(\phi \to \phi_H)$. We show the numerical results\footnote{In order to solve Eq.\ \eqref{eq:zetaedo} for the perturbation $\psi$ we used a part of numerical code created by the authors of \cite{prescription-justification-2}, which is available at \url{https://www.princeton.edu/physics/research/high-energy-theory/gubser-group/code-repository/}.} for $\zeta/s$ in Fig.\ \ref{fig:zeta}. The bulk viscosity displays a peak near the phase transition but its magnitude is still smaller than $\eta/s$. 

For completeness, we also provide a fit to the numerical results of Fig.\ \ref{fig:zeta}. The form of the function suggests a resonance-like fitting function. With this in mind we used a five-parameter trial function
\begin{equation}
\label{eq:zetafit}
\frac{\zeta}{s}\left(x=\frac{T}{T_c}\right) = \frac{a}{\sqrt{\left(x-b\right)^2+c^2}} + \frac{d}{x^2+e^2},
\end{equation}
where $a$ to $e$ are fit parameters (and $T_c=143.8$ MeV). The first term of Eq.\ \eqref{eq:zetafit} describes the resonance-like peak of Fig.\ \ref{fig:zeta} while the second term describes a smooth background away from the peak. Using the parameters in Table \ref{tab:parameterszeta}, we obtain the fit shown in Fig.\ \ref{fig:zeta}, which gives a good description of our numerical results. 
\begin{table}[htp]
\caption{Parameters for the fit of $\zeta/s$ using Eq.\ \eqref{eq:zetafit}. }
\begin{center}
\begin{tabular}{cccccc}

\hline
\hline

$a$ & $b$ & $c$ & $d$ & $e$ \\ \hline
0.01162 & 1.104 & 0.2387 & -0.1081 & 4.870 \\ \hline\hline

\end{tabular}
\end{center}
\label{tab:parameterszeta}
\end{table}

At this point we have then directly computed 6 transport coefficients (besides matching lattice QCD thermodynamics): $\eta/s$, $\zeta/s$, $\tau_\pi$, $\kappa$, $\kappa^*$, and $\xi_5$. Among these coefficients, only $\eta/s$ was found to be a constant with $T$ - all the other coefficients displayed some nontrivial behavior near the crossover phase transition. 

Next, we use these results to give our best estimate for the temperature dependence of 6 other coefficients: $\lambda_3$, $\lambda_4$, $\xi_3$, $\xi_4$, $\xi_6$, and $\tau_\Pi$.

\subsubsection*{Estimates for the coefficients $\xi_3$, $\xi_4$, $\xi_6$, $\lambda_3$, and $\lambda_4$}

Let us now examine three other second order transport coefficients of non-conformal hydrodynamics, $\xi_3$, $\xi_4$, and $\xi_6$, which satisfy the following constraints \cite{moore}
\begin{align}
\xi_6 & = c_s^2 \left( 3 T \frac{d \kappa}{dT} -2T \frac{d\kappa^*}{dT} + 2 \kappa^* - 3 \kappa \right) - \kappa + \frac{4\kappa^*}{3} + \frac{\lambda_4}{c_s^2}, \label{eq:xi62} \\ 
\xi_3 & = \frac{3 c_s^2}{2} T \left(\frac{d \kappa^*}{dT} -  \frac{d \kappa}{dT} \right) + \frac{3}{2} (c_s^2-1) \left( \kappa^* - \kappa\right) - \frac{\lambda_4}{c_s^2} + \frac{1}{4} \left( c_s^2 T \frac{d \lambda_3}{dT} - 3 c_s^2 \lambda_3 + \frac{\lambda_3}{3} \right), \label{eq:xi32} \\
\xi_4 & = -\frac{\lambda_4}{6} -\frac{c_s^2}{2} \left( \lambda_4 + T \frac{d\lambda_4}{dT} \right) + c_s^4 (1-3c_s^2) \left( T \frac{d \kappa}{dT} - T \frac{d \kappa^*}{dT} + \kappa^* - \kappa \right) + \nonumber \\
& - c_s^6 T^3 \frac{d^2}{dT^2} \left( \frac{\kappa - \kappa^*}{T} \right), \label{eq:xi42}
\end{align}
where the second order coefficients $\lambda_3$ and $\lambda_4$ are given by the following Kubo's formulas involving Euclidean 3-point functions \cite{moore,Moore:2010bu}
\begin{align}
\lambda_3 & = 2 \kappa^* -4 \lim_{p_z, q_z \to 0} \frac{\partial^2}{\partial p_z\partial q_z} G_E^{xt,yt,xy} (p_t=0, \vec{p}, q_t=0, \vec{q}) \label{eq:lambda3}, \\
\lambda_4 & = -2 \kappa^* + \kappa - \frac{c_s^4}{2} \lim_{p_x, q_y \to 0} \frac{\partial^2}{\partial p_x\partial q_y} G_E^{tt,tt,xy} (p_t=0, \vec{p}, q_t=0, \vec{q})\,. \label{eq:lambda4}
\end{align}
In order to compute $\xi_3$, $\xi_4$, and $\xi_6$ using the constraints \eqref{eq:xi62} to \eqref{eq:xi42} it is necessary to evaluate $\lambda_3$ and $\lambda_4$. However, the holographic computation of 3-point functions is far more involved than the computation of 2-point functions. In fact, for the strongly coupled SYM plasma $\lambda_3$ has been evaluated explicitly by computing the Euclidean 3-point function $G_E^{xt,yt,xy}(p,q)$, yielding $\lambda_3 = 0$ \cite{Arnold:2011ja}. Since $\kappa^* = 0$ in a conformal theory, from (\ref{eq:lambda3}) we obtain that for a strongly coupled SYM
\begin{equation}
\label{eq:3point3}
\lim_{p_z, q_z \to 0} \frac{\partial^2}{\partial p_z\partial q_z} G_E^{xt,yt,xy} (p_t=0, \vec{p}, q_t=0, \vec{q}) = 0.
\end{equation}
In order to evaluate $\lambda_4$ one should in principle compute the Euclidean 3-point function $G_E^{tt,tt,xy}(p,q)$. However, there is a shortcut which makes use of the constraint \eqref{eq:xi62}. In a 4-dimensional CFT, we know that $c_s^2 = 1/3$ and we also know that $\kappa^* = 0$ and that $\xi_{3,4,5,6} = 0$, since these are coefficients of non-conformal hydrodynamics. Then, from Eq.\ \eqref{eq:xi62} we deduce that in a CFT $\lambda_4 = 0$, which agrees with \cite{romatschke}. Then, from Eq.\ \eqref{eq:lambda4}, we conclude that in a strongly coupled CFT
\begin{equation}
\label{eq:3point1}
\lim_{p_x, q_y \to 0} \frac{\partial^2}{\partial p_x\partial q_y} G_E^{tt,tt,xy} (p_t=0, \vec{p}, q_t=0, \vec{q}) = \frac{2 \kappa}{c_s^4}.
\end{equation}

The evaluation of the full 3-point functions required to compute the coefficients $\lambda_3$ and $\lambda_4$ in the effective model of Einstein+Scalar gravity where the metric is only known numerically is far beyond the scope of this work. In order to fully determine them it is necessary to compute the full bulk-to-boundary and bulk-to-bulk propagators - and it is very difficult to compute these functions in terms of a numerical metric such as the one used in this work.

\begin{figure}
\centering
  \includegraphics[width=.5\linewidth]{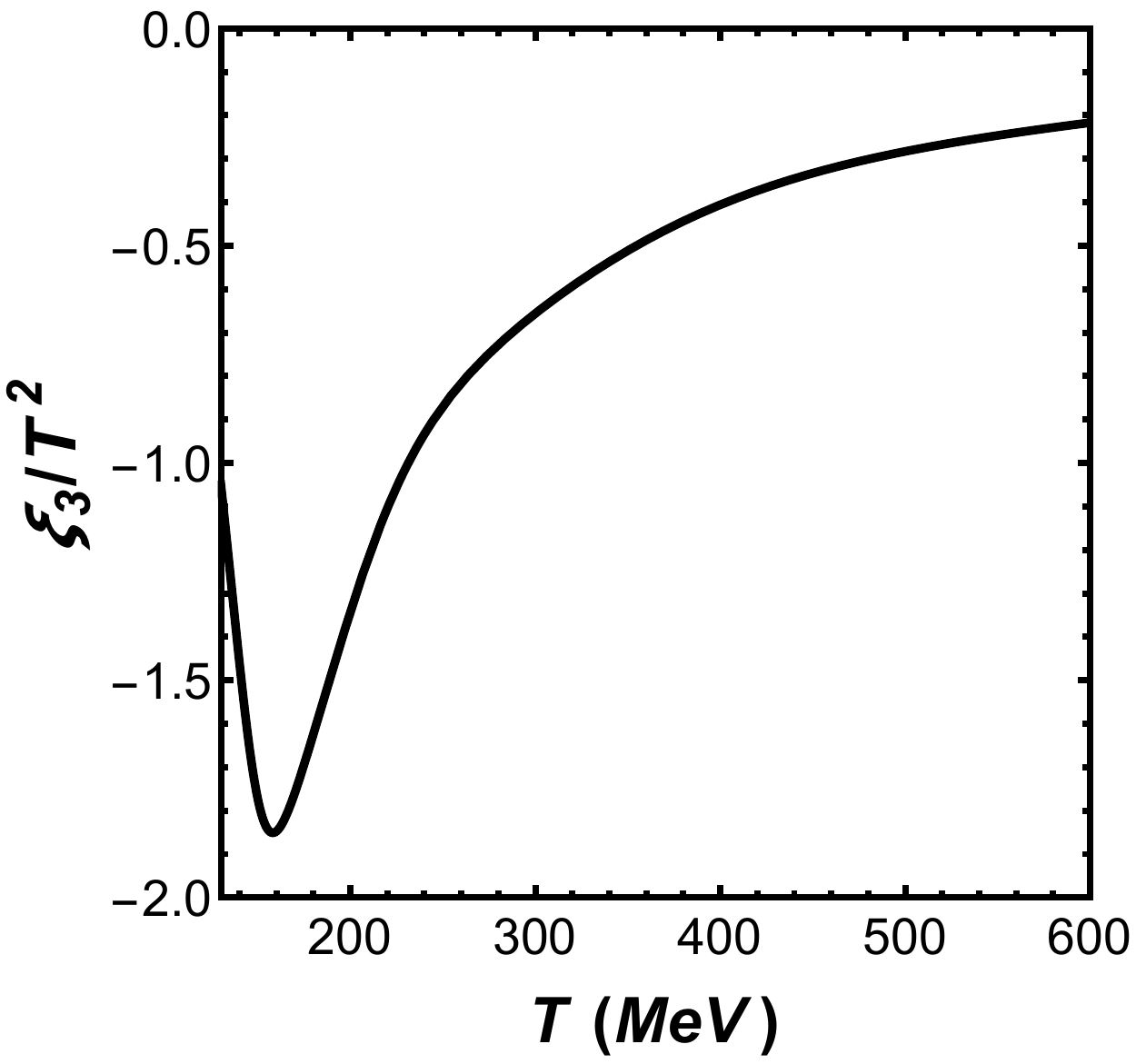}
  \caption{$\xi_3/T^2$ as a function of the temperature $T$ for the bottom-up holographic model.}
  \label{fig:xi3}      
\end{figure}

\begin{figure}
\centering
  \includegraphics[width=.5\linewidth]{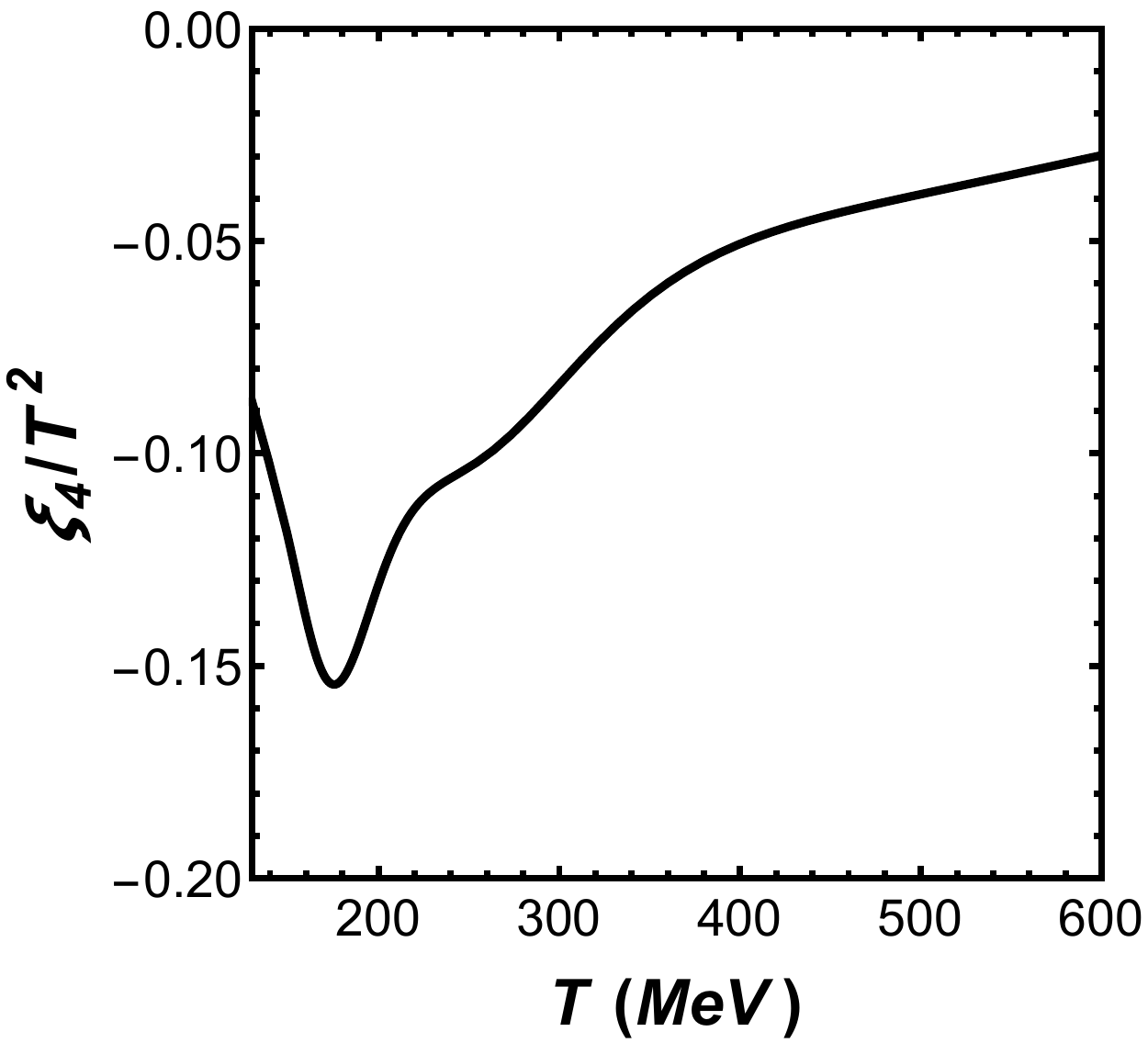}
  \caption{$\xi_4/T^2$ as a function of the temperature $T$ for the bottom-up holographic model.}
  \label{fig:xi4}      
\end{figure}

\begin{figure}
\centering
  \includegraphics[width=.5\linewidth]{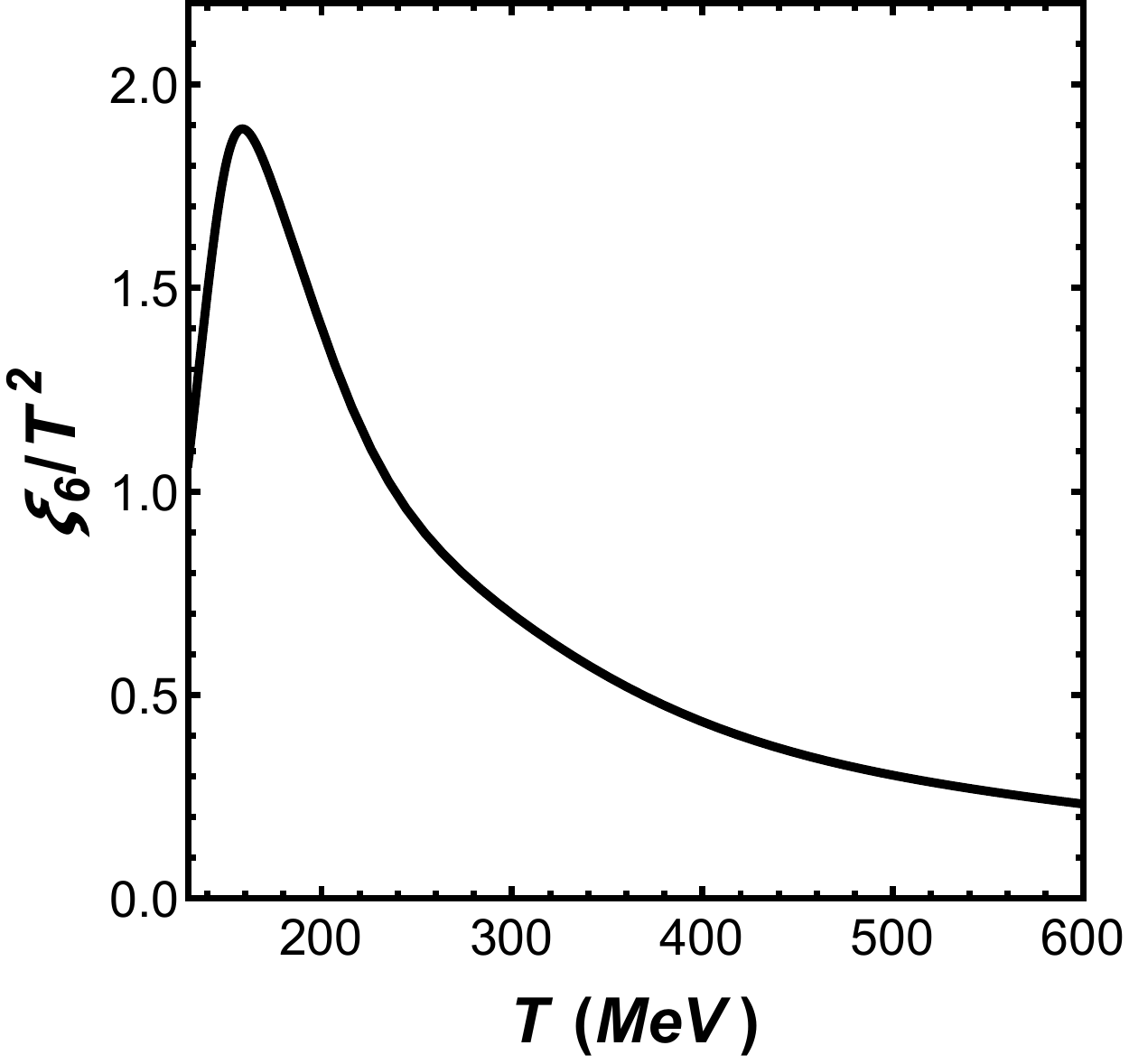}
  \caption{$\xi_6/T^2$ as a function of the temperature $T$ for the bottom-up holographic model.}
  \label{fig:xi6}
\end{figure}

Thus, in this paper we will resort to a sort of ``hybrid CFT/non-CFT" approximation. For the evaluation of the second order coefficients $\lambda_3$ and $\lambda_4$ from Eqs.\ \eqref{eq:lambda3} and \eqref{eq:lambda4} we shall mix the CFT 3-point functions \eqref{eq:3point3} and \eqref{eq:3point1} with the full non-conformal results for $\kappa$ and $\kappa^*$. In this case, the resulting approximations for $\lambda_3$ and $\lambda_4$ give
\begin{equation}
\lambda_3 = - \lambda_4 = 2 \kappa^*\,.
\end{equation}
With these approximations, and the full non-conformal results for $\kappa$, $\kappa^*$, and $c_s^2$, we can approximately evaluate $\xi_3$, $\xi_4$, and $\xi_6$ from Eqs.\ \eqref{eq:xi62} to \eqref{eq:xi42}. This sort of approximation provides our best estimate for these coefficients given the current lack of knowledge about 3-point functions in non-conformal holographic plasmas that display a crossover phase transition. In Figs.\ \ref{fig:xi3} to \ref{fig:xi6} we show the results for $\xi_6/T^2$, $\xi_3/T^2$, and $\xi_4/T^2$ as functions of $T$ - all of these coefficients vary rapidly near the phase transition. We note that the approximations done here for these coefficients constitute the first deviations from the ultraviolet conformal regime and, therefore, they are much more reliable at high temperatures.

\subsubsection*{A lower bound estimate for $\tau_{\Pi}$}

\begin{figure}
\centering
  \includegraphics[width=.5\linewidth]{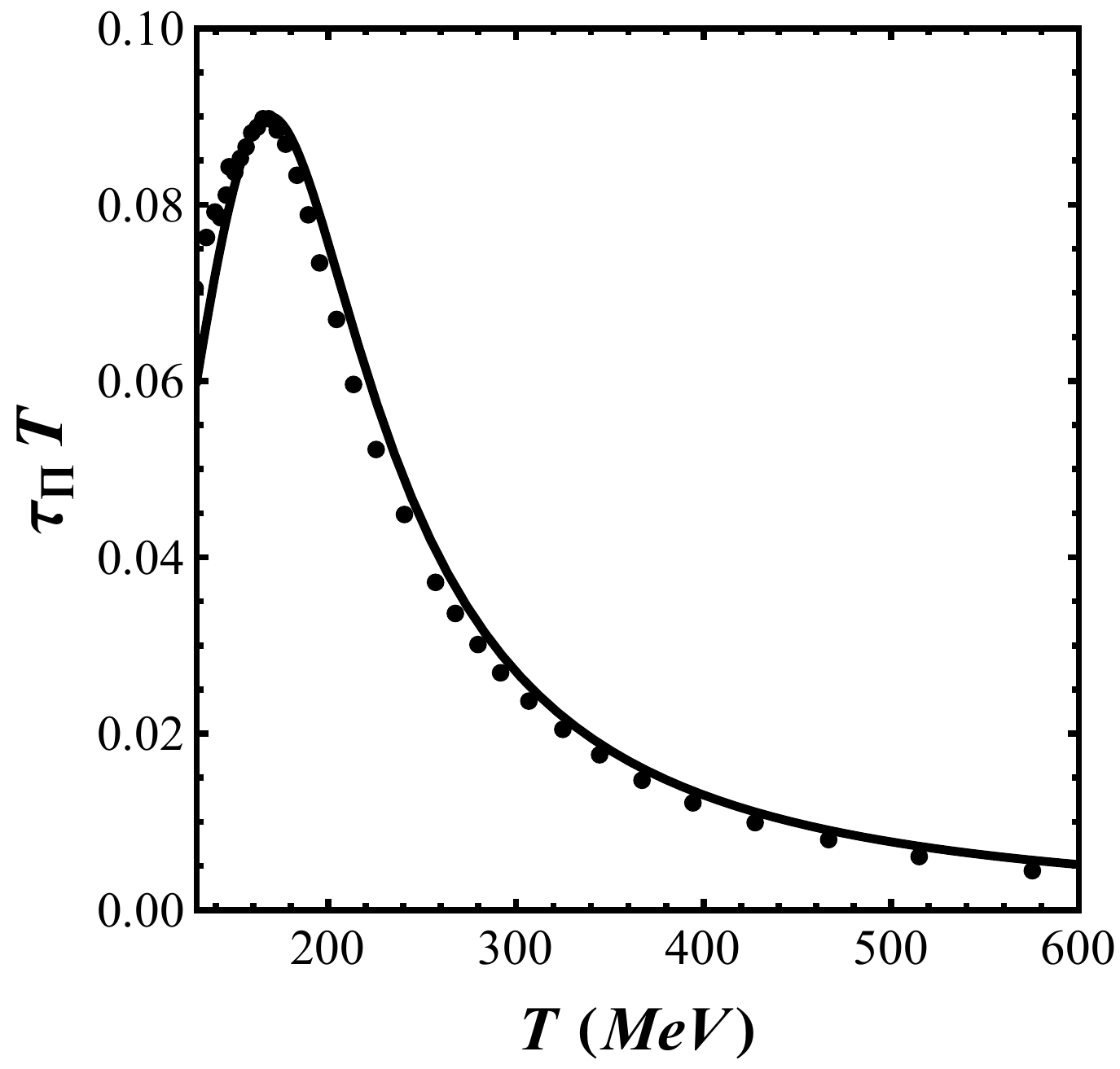}
  \caption{A lower bound for $\tau_{\Pi} T$ as a function of the temperature $T$ for the bottom-up holographic model. The black points correspond to our numerical results while the black curve is the fit in Eq.\ \eqref{eq:tauPIfit} with the parameters in Table \ref{tab:parameterstauPI}.}
  \label{fig:taupibulk}      
\end{figure}

Ref.\ \cite{Pu:2009fj} derived, using the asymptotic causality condition, a relation among the transport coefficients $\tau_\pi$, $\tau_\Pi$, $\eta$, and $\zeta$ 
\begin{equation}
\frac{\zeta}{s \tau_{\Pi} T} + \frac{\eta}{s \tau_{\pi} T} \leq 1 - c_s^2\,.
\label{conditiontauPi}
\end{equation}
The computation of the transport coefficient $\tau_\Pi$ directly from its retarded Green's function, as was done for the shear coefficient $\tau_\pi$, is beyond the scope of this paper. However, we note that one can use the relation (\ref{conditiontauPi}) to obtain a lower bound for the coefficient $\tau_{\Pi}$ that can still be useful for hydrodynamic modeling of the QGP \footnote{A similar idea was used in \cite{NoronhaHostler:2012ug} to estimate the coefficient $\tau_\pi$ in a hadronic gas with Hagedorn resonances given the result found in this case for $\eta/s$ \cite{NoronhaHostler:2008ju}.}. The result for this lower bound in our holographic model is shown in Fig.\ \ref{fig:taupibulk} (this was computed using directly the fitting functions and Eq.\ (\ref{conditiontauPi})). This is the smallest value of $\tau_{\Pi} T$ in our model that is still consistent with causality and linear stability \cite{Pu:2009fj}. One can see that this coefficient displays a peak near the phase transition, as was the case for $\tau_\pi T$, but it becomes very small at high temperatures, as expected due to conformal invariance. The results in Fig.\ \ref{fig:taupibulk} also admit a fit using the following parametrization, 
\begin{equation}
\label{eq:tauPIfit}
\tau_{\Pi} T \left(x=\frac{T}{T_c}\right) = \frac{a}{\sqrt{\left(x-b\right)^2+c^2}} + \frac{d}{x},
\end{equation}
with the corresponding fit parameters $a$ to $d$ being given in Tab. \ref{tab:parameterstauPI}.

\begin{table}[htp]
\caption{Parameters for the fit of $\tau_{\Pi}T$ using Eq.\ \eqref{eq:tauPIfit}. }
\begin{center}
\begin{tabular}{cccccc}

\hline
\hline

$a$ & $b$ & $c$ & $d$ \\ \hline
0.05298 & 1.131 & 0.3958 & -0.05060 \\ \hline\hline

\end{tabular}
\end{center}
\label{tab:parameterstauPI}
\end{table}

Therefore, in this section we presented results (computed within different levels of approximations) for 12 transport coefficients that appear at second order in the gradient expansion of a non-conformal plasma that has thermodynamic properties similar to those found for the QGP on the lattice: $\eta/s$, $\zeta/s$, $\tau_\pi$, $\kappa$, $\kappa^{*}$, $\xi_5$ as well as $\xi_3$, $\xi_4$, $\xi_6$, $\lambda_3$, $\lambda_4$, and $\tau_\Pi$. However, the equations of motion obtained from (\ref{defineshearstress}) and (\ref{definebulk}) are not in a form suitable for numerical implementation. In the next section we use the gradient expansion to find another 2nd order theory, similar to Israel-Stewart theory\cite{ist}, that can be readily used in phenomenological studies of the QGP hydrodynamical evolution in heavy ion collisions.

\section{Israel-Stewart-like 2nd order hydrodynamics for a non-conformal relativistic fluid}\label{ISsection}

It is known that relativistic NS theory leads to acausal propagation of sound and shear linear disturbances around an equilibrium state at rest and that in the case of a moving background fluid these disturbances are unstable \cite{acausal,shipu1}. It can be shown that the 2nd order theory in Eqs.\ (\ref{defineshearstress}) and (\ref{definebulk}) is linearly unstable above a certain critical wavenumber even for a fluid at rest, as demonstrated in Appendix \ref{appendixhugo} (see also \cite{hugofuturo}). Note also that the inclusion of these spatial gradients cannot modify the theory at very large momenta $k$ where the asymptotic causality conditions are defined \cite{Pu:2009fj}. However, it is not the purpose of hydrodynamics to accurately describe small wavelength phenomena - this is beyond the scope of this effective theory. Nevertheless, it is desirable that for a system that may be coupled to gravity causality (and stability!) are preserved. 

In this section we use the 2nd order theory in (\ref{defineshearstress}) and (\ref{definebulk}) to construct a relaxation-type theory (in curved spacetime) that is similar to that considered by Israel and Stewart \cite{ist} and also to those that appear naturally within kinetic theory using the moments method \cite{Denicol:2012cn}. The main idea, already pursued in \cite{brsss} in the case of a conformal fluid, is to write a simple relaxation-type theory that reduces to the gradient expansion theory in (\ref{defineshearstress}) and (\ref{definebulk}) in its asymptotic hydrodynamical limit. In this section we shall use the same procedure generalized to the case of a non-conformal fluid.

First, the terms involving comoving derivatives of $\sigma^{\mu\nu}$ and $\theta$ on the right-hand side of (\ref{defineshearstress}) and (\ref{definebulk}) are transferred to the left-hand side of those equations using the lowest order substitutions $\sigma^{\mu\nu} \to - \pi^{\mu\nu}/\eta$ and $\theta \to -\Pi/\zeta$. Moreover, we use the leading order expression $D s = -s \theta + \ldots$ in \eqref{entropyproduction} to simplify some of the terms (note that we are neglecting terms of third order in gradients that would appear in this general procedure). The same leading order substitution is done in the remaining 2nd order terms on the right-hand side of the equations with the exception of the term $\sim \tau_\pi\eta\, \theta \sigma^{\mu\nu}$ in \eqref{defineshearstress} where only the substitution $\sigma^{\mu\nu} \to -\pi^{\mu\nu}/\eta$ is done. This choice is made to make it explicit that the combination $D\pi^{\langle\mu\nu \rangle}+4\theta \pi^{\mu\nu}/3$ is the correct combination that survives in the conformal limit \cite{brsss} (being, thus, homogeneous under Weyl transformations). One can then show that this leads to 
\bea
\tau_\pi \left(D\pi^{\langle \mu\nu\rangle}+\frac{4\theta}{3}\pi^{\mu\nu}\right) +\pi^{\mu\nu} &=& -\eta \sigma^{\mu\nu} + \kappa \left(\mathcal{R}^{\langle \mu\nu\rangle}-2u_\alpha u_\beta \mathcal{R}^{\alpha \langle \mu\nu \rangle \beta}\right) +\tau_\pi\, \pi^{\mu\nu}\,\,D \ln \left(\frac{\eta}{s}\right)    \nonumber \\ &+& \frac{\lambda_1}{\eta^2}  \pi_\lambda^{\langle \mu} \pi^{\nu\rangle\lambda}- \frac{\lambda_2}{\eta} \pi_\lambda^{\langle \mu} \Omega^{\nu\rangle\lambda} -\lambda_3  \Omega_\lambda^{\langle \mu} \Omega^{\nu\rangle\lambda} +2 \kappa^{*} \,u_\alpha u_\beta \mathcal{R}^{\alpha \langle \mu\nu \rangle \beta} \nonumber \\ &+& \tau_\pi^{*}\pi^{\mu\nu}\,\frac{\Pi}{3\zeta}+\lambda_4 \nabla^{\langle \mu}\ln s \,\nabla^{\nu\rangle} \ln s\,
\label{definenewshearstress}
\eea
and
\bea
\tau_\Pi \left(D\Pi+\Pi \theta\right)+\Pi  &=& -\zeta \theta  +\frac{\xi_1}{\eta^2} \pi_{\mu\nu}\pi^{\mu\nu} +\frac{\xi_2}{\zeta^2}\Pi^2 +\tau_\Pi\,\Pi\,\,D \ln \left(\frac{\zeta}{s}\right) \nonumber \\ &+& \xi_3 \Omega_{\mu\nu}\Omega^{\mu\nu}+\xi_4 \nabla_\mu^{\perp} \ln s\,\nabla_\perp^\mu \ln s+\xi_5 \mathcal{R} + \xi_6 u^\mu u^\nu \mathcal{R}_{\mu\nu}\,.
\label{definenewbulk}
\eea
These are nonlinear, coupled partial differential equations of relaxation-type for the new dynamical variables $\pi^{\mu\nu}$ and $\Pi$ in curved spacetime that require (independent) initial conditions in order to solve Eqs.\ (\ref{definenewshearstress}) and (\ref{definenewbulk}) together with the conservation equations (\ref{conservationeqs})\footnote{This theory is qualitatively different than that in the gradient expansion - the dissipative parts of the energy-momentum tensor have their own differential equations and, thus, its initial conditions are not determined by the initial conditions for the hydrodynamic variables $\varepsilon$ and $u^\mu$.}.  These equations are similar to those found in Israel-Stewart theory \cite{ist} and they possess most of the terms found in kinetic theory in flat spacetime \cite{Denicol:2012cn}\footnote{Ref.\ \cite{Denicol:2012cn} used the Boltzmann equation and a completely distinct power-counting scheme to deal with the gradients in comparison to the one used to derived our equations for $\pi^{\mu\nu}$ and $\Pi$. For instance, according to the power-scheme of \cite{Denicol:2012cn}, the vast majority of their terms of order $\mathcal{O}(K_n^2)$ do not appear in our equations. Thus, perfect agreement among these approaches should not really be expected.}. Furthermore, note that the asymptotic, leading order solution of these equations necessarily reduce to the gradient expansion in Eqs.\ (\ref{defineshearstress}) and (\ref{definebulk}) up to $\mathcal{O}(K_n^2)$. 

However, for phenomenological applications in heavy ion collisions, the terms containing spacetime curvatures are negligible and can, thus, be dropped. Moreover, note that while in our holographic model $\eta/s$ is a constant, in general one should keep the term involving $D\,\ln(\eta/s)$ to make sure that this theory reduces to the correct gradient expansion theory if the asymptotic limit $\pi^{\mu\nu} \to -\eta \sigma^{\mu\nu}$ is taken. For the same reason, one should keep $D\,\ln(\zeta/s)$ and, in fact, in our model since $\zeta/s$ is a not a constant its comoving derivative is not zero (note also that the conservation equations imply that, to lowest order, $D\,\ln(\eta/s)\,, D\,\ln(\zeta/s) \sim -\theta$).

This leads us to the following (reduced) set of equations that can be used in hydrodynamic simulations of the QGP
\bea
\tau_\pi \left(D\pi^{\langle \mu\nu\rangle}+\frac{4\theta}{3}\pi^{\mu\nu}\right) +\pi^{\mu\nu} &=& -\eta \sigma^{\mu\nu} +\frac{\lambda_1}{\eta^2}  \pi_\lambda^{\langle \mu} \pi^{\nu\rangle\lambda}- \frac{\lambda_2}{\eta} \pi_\lambda^{\langle \mu} \Omega^{\nu\rangle\lambda} - \lambda_3  \Omega_\lambda^{\langle \mu} \Omega^{\nu\rangle\lambda} \nonumber \\ &+& \tau_\pi\, \pi^{\mu\nu}\,D \,\ln\left(\frac{\eta}{s}\right) +\tau_\pi^{*}\pi^{\mu\nu}\,\frac{\Pi}{3\zeta} +\lambda_4 \nabla^{\langle \mu}\ln s \,\nabla^{\nu\rangle} \ln s\,
\label{definenewshearstress1}
\eea
and
\bea
\tau_\Pi \left(D\Pi+\Pi \theta\right)+\Pi  &=& -\zeta \theta  +\frac{\xi_1}{\eta^2} \pi_{\mu\nu}\pi^{\mu\nu} +\frac{\xi_2}{\zeta^2}\Pi^2 + \xi_3 \Omega_{\mu\nu}\Omega^{\mu\nu} \nonumber \\ &+& \tau_\Pi\,\Pi\,D\,\ln\left(\frac{\zeta}{s}\right)+\xi_4 \nabla_\mu^{\perp} \ln s\,\nabla_\perp^\mu \ln s\,.
\label{definenewbulk1}
\eea
One can show that the hydrodynamic theory described above is linearly stable and causal according to the criteria of \cite{Pu:2009fj} and, thus, it should be suitable for implementation in modern numerical viscous hydrodynamic codes such as \cite{Schenke:2010rr,Bozek:2011ua,Noronha-Hostler:2013gga,DelZanna:2013eua,Karpenko:2013wva,Shen:2014vra}. We stress that the terms involving $D\ln(\eta/s)$ and $D\ln(\zeta/s)$ in the equations above are needed to recover the correct 2nd order gradient expansion and should not in principle be neglected in numerical simulations. 

Among the 13 transport coefficients left in the equations above, results for 8 of them have already been presented in this paper while we have not yet discussed the coefficients $\lambda_1$, $\lambda_2$, $\xi_1$, $\xi_2$, and $\tau_\pi^*$. The coefficients $\lambda_1$ and $\lambda_2$ have been studied at weak coupling in \cite{York:2008rr} and at strong coupling in \cite{brsss,Bhattacharyya:2008jc,Arnold:2011ja}. The SYM values of these coefficients computed at strong coupling via holography are\footnote{For CFT's with a holographic description involving two derivatives in the bulk it was found in Refs.\ \cite{Erdmenger:2008rm,Haack:2008xx} that $4\lambda_1+\lambda_2=2\eta \tau_\pi$ (see also \cite{Haehl:2014zda}). Moreover, Ref.\ \cite{Grozdanov:2014kva} has recently found that this relation remains valid in a SYM plasma even when the leading order finite t'Hooft coupling corrections are taken into account.} 
\begin{equation}
\lambda_1 = 2 \frac{\eta^2}{s T}\,,\qquad \lambda_2 = -\ln 2 \frac{\eta}{\pi T}\,.
\label{lambdaSYM}
\end{equation}
We are not aware of any calculation of these coefficients in a non-conformal strongly coupled plasma with a crossover transition. However, within the phenomenological ``spirit" of this section and since we currently lack a better way to compute them, one may take the SYM expressions above for the non-conformal case at hand. This would imply that $\lambda_1/T^2 \sim s/T^3$ and $\lambda_2/T^2 \sim - s/T^3$. Thus, in this case these coefficients would display the same sharp rise observed by the entropy density near the phase transition.

Knowledge about the coefficients $\xi_1$ and $\xi_2$ is much more scarce. These coefficients only appear in non-conformal fluids and very little is known about them at strong coupling. An exception is the strongly coupled non-conformal plasma studied in Ref.\ \cite{Kanitscheider:2009as} constructed via dimensional reduction of a higher dimensional pure gravity action. In this case, these coefficients can be extracted using the fluid/gravity correspondence and they read \cite{romatschke} \footnote{The transport coefficients in the theory \cite{Kanitscheider:2009as} are known analytically and, even though their numerical values are different than the ones found in this paper (their theory is different than ours), qualitatively they possess the same features found here - $\xi_5$, $\xi_6$ have the same signal as ours and would also display a peak where $c_s^2$ has a minimum. However, their $\tau_\pi T$ is a constant while ours has a peak near the phase transition.}
\be
\xi_1 =\lambda_1\left(\frac{1}{3}-c_s^2\right)\,, \qquad  \xi_2 = 2\eta \tau_\pi c_s^2\left(\frac{1}{3}-c_s^2\right)\,.
\label{x1x2}
\ee
Also, in this theory one finds \cite{romatschke}
\be
\tau_\pi^{*} = -3\tau_\pi \left(\frac{1}{3}-c_s^2\right)\,. 
\label{taupiestrela}
\ee
In the absence of a better estimate for these three coefficients above in the non-conformal plasma proposed in this paper, it may be useful in hydrodynamic simulations of the QGP to use the expressions in (\ref{x1x2}) and (\ref{taupiestrela}) hoping that they get at least part of the non-conformal dynamics near the phase transition. Notice, however, that these expressions contain $(\frac{1}{3} -c_s^2)$ and that this is an ubiquitous factor in non-conformal transport coefficients - for instance, the bulk viscosity of our model is proportional to this factor \cite{Gubser:2008yx,Buchel:2007mf}. Thus, it is reasonable to assume that these expressions may describe the temperature behavior of these coefficients in our model as well. Moreover, we would like to remark that our calculation for the transport coefficients are qualitatively consistent with the results found in \cite{Bigazzi:2010ku} for the large $T$ expansion of the 2nd order transport coefficients for the non-conformal plasma dual to the Chamblin-Reall background \cite{Chamblin:1999ya}.

Therefore, after this long discussion, the hydrodynamic theory described by Eqs.\ (\ref{definenewshearstress1}) and (\ref{definenewbulk1}) together with the corresponding conservation equations (\ref{conservationeqs}) may be a good starting point for phenomenological applications of relativistic non-conformal hydrodynamics for the strongly-coupled QGP formed in heavy ion collisions\footnote{Note that the thermodynamics of the model is very similar to lattice data and, thus, in hydrodynamic simulations one may as well just use directly the lattice data for the thermodynamical quantities such as $c_s^2$ or $s$.}. To facilitate the use of our results in hydrodynamic simulations, we have provided a guide for all the relevant formulas for the 13 transport coefficients of this second order theory in Appendix \ref{summaryhydro}.

\section{Conclusions}\label{concsec}

In this paper we used the gauge/gravity duality to determine the transport coefficients of a non-conformal, strongly interacting non-Abelian plasma that displays a crossover transition similar to that found for the QGP determined via lattice calculations. The 5-dimensional gravity dual model involves the metric coupled with a dynamical scalar field and its simplicity and capability of describing several nontrivial features of the QGP have motivated us to pursue the calculations of the several transport coefficients shown in this paper. 

We first obtained holographic formulas for the transport coefficients $\kappa$ and $\tau_\pi$ present in the second-order gradient expansion of relativistic hydrodynamics in curved spacetime. Our method to compute these coefficients could also be applied in the case where the gravity dual possesses other fields besides the scalar field, such as the case of an Einstein+Scalar+Maxwell model with at most two derivatives in the action. Also, besides the well-known result for $\eta/s$, we were also able to directly compute five other coefficients that appear at second-order in the derivative expansion: $\tau_\pi$, $\zeta/s$, $\kappa$, $\kappa^*$, and $\xi_5$. Apart from $\eta/s$, all of these coefficients displayed nontrivial behavior near the crossover transition. In particular, $\tau_\pi T$, $\zeta/s$, and $\xi_5/T^2$ display a peak near the transition while $\kappa^*/T^2$ is similar to $c_s^2$ (though it is negative) in that it displays a minimum at the crossover transition. On the other hand, $\kappa/T^2$ rises monotonically with $T$ until it reaches its conformal limit (i.e., its value in SYM). Our values for $\tau_\pi T$ only deviates from the SYM result $(2-\ln 2)/(2\pi)$ at low temperatures. The coefficients $\kappa$, $\kappa^*$, and $\xi_5$ only contribute directly to the equations of motion in a curved spacetime.

\begin{figure}
\centering
  \includegraphics[width=.5\linewidth]{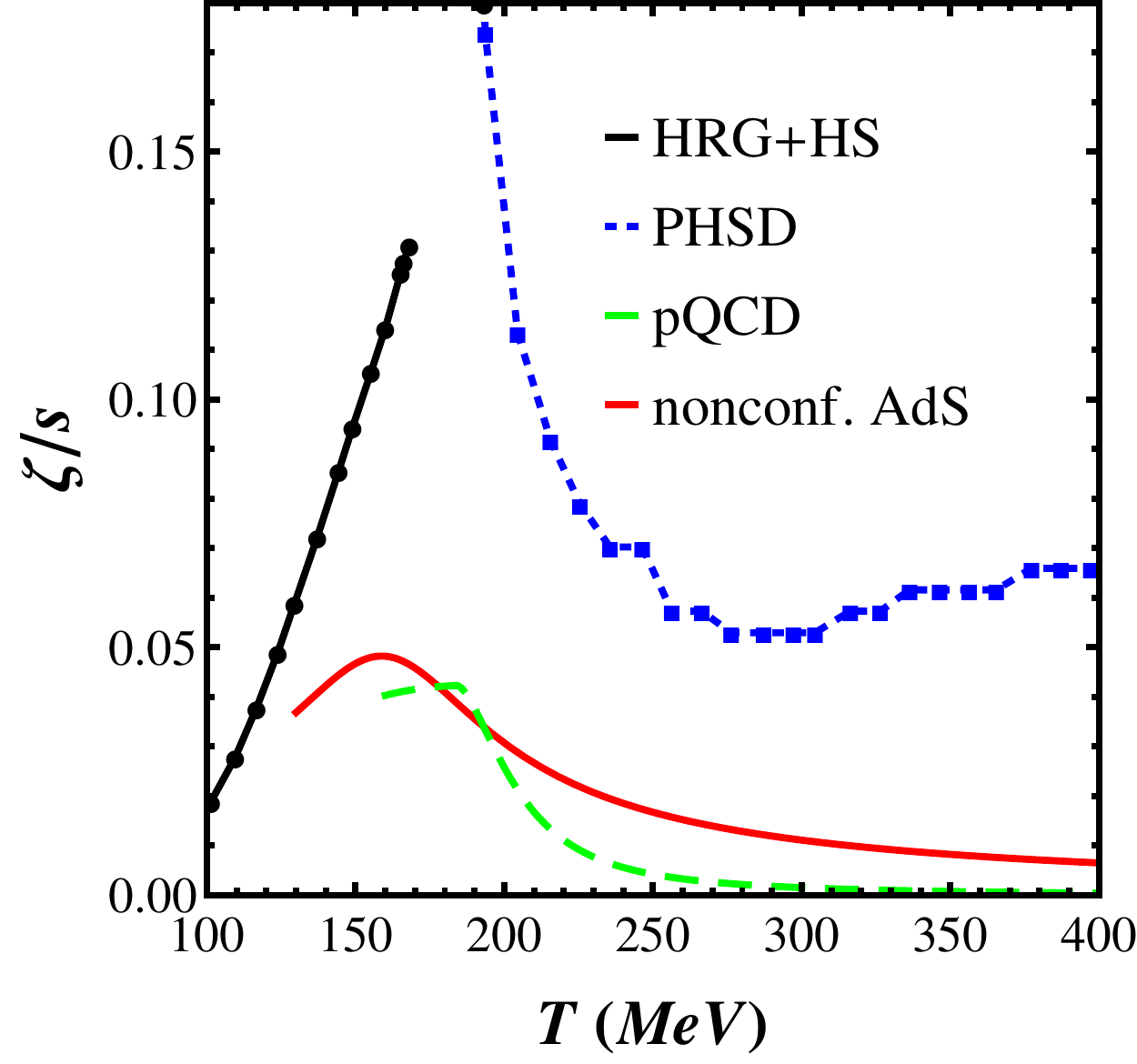}
  \caption{(Color online) Comparison between different calculations of $\zeta/s$. The solid red curve shows the holographic calculation of this paper (cut at $T=130$ MeV), the dashed green curve shows an extrapolation of the pQCD results of \cite{Arnold:2006fz} towards low temperatures, the blue squares shows the calculation using the PHSD model \cite{Ozvenchuk:2012kh}, while the black points shows the hadronic calculation from \cite{Kadam:2014cua}.}
  \label{newplotzeta}      
\end{figure}

Our $\zeta/s$ is in general smaller than that found in other works \cite{Ozvenchuk:2012kh,Kadam:2014cua} as it is clear in\footnote{We thank J.~Noronha-Hostler for making this plot available to us.} Fig.\ \ref{newplotzeta}, though it is similar in magnitude to the pQCD calculation in Ref.\ \cite{Arnold:2006fz}. Thus, at least according to our calculations, cavitation \cite{Torrieri:2007fb,Torrieri:2008ip,Rajagopal:2009yw} induced by a large $\zeta/s$ in the phase transition is not likely to occur in the QGP (a similar conclusion was reached in \cite{Jaiswal:2013fc} using a kinetic theory-derived bulk relaxation time in a Bjorken expanding fluid). However, it could be that the other coefficients that appear in the bulk equation, together with the shear-bulk coupling terms such as the $\pi^{\mu\nu}\Pi$ term in (\ref{definenewshearstress1}), may in the end take the evolving plasma towards cavitation. This is an interesting possibility that can be checked in numerical simulations. 

We used these calculations to provide estimates for the other coefficients $\xi_3$, $\xi_4$, $\xi_6$, $\lambda_3$, $\lambda_4$, and $\tau_\Pi$. We found that $\xi_3/T^2$ and $\xi_4/T^2$ are negative and have a minimum near the transition while $\xi_6/T^2$ is positive and displays a peak. Note that $\xi_6$ is only relevant in curved spacetime while $\xi_3$, $\lambda_3$, $\xi_4$, $\lambda_4$ do affect the motion of the fluid in flat spacetime. In fact, $\xi_3$ and $\lambda_3$ are related to the vorticity tensor $\Omega_{\mu\nu}$ whose role in hydrodynamic simulations has not yet been investigated in detail\footnote{Note that for (0+1) purely Bjorken hydrodynamics this term disappears even in a non-conformal plasma.}. Moreover, $\xi_4$ and $\lambda_4$ also have not been investigated in hydrodynamic calculations and, thus, we hope the results of this paper may serve as motivation for a detailed investigation of their effects. In this paper we have used the asymptotic causality condition \cite{Pu:2009fj} to obtain the lowest possible value of the coefficient $\tau_\Pi$ associated with bulk viscosity relaxation. In this case, this lower bound for $\tau_\Pi T$ displays a peak near the transition (though its value at the peak is relatively small, in agreement with the small value of $\zeta/s$ found here)\footnote{We remark that the coefficients $\tau_\pi$ and $\tau_\Pi$, as defined via the gradient expansion, do not necessarily correspond to relaxation time coefficients. Clearly, relaxation time coefficients require at least relaxation equations for $\pi_{\mu\nu}$ and $\Pi$, such as those in Israel-Stewart theory. In fact, according to their definitions in the gradient expansion, these coefficients do not even need to be positive. For instance, \cite{Schaefer:2014aia} has found an example of a gravity dual in which $\tau_\pi < 0$ as defined via the gradient expansion. This, however, was shown to not generate instabilities as it would have been the case if that coefficient were indeed a measure of shear relaxation. As discussed in \cite{Denicol:2011fa}, only the coefficients extracted from the poles of retarded correlators do have the meaning of shear or bulk relaxation time coefficients, which is not generally the case for the coefficients defined via derivatives of retarded Green's functions (such as in the gradient expansion).}.

We have used the 2nd-order gradient expansion equations to construct an Israel-Stewart-like theory in flat spacetime, shown in Eqs.\ (\ref{definenewshearstress1}) and (\ref{definenewbulk1}), which gives equations of motion that preserve causality and are linearly stable around thermal equilibrium\footnote{It is important to remark that the procedure used here to find relaxation equations using the gradient expansion has some ambiguities. Clearly, the set of relaxation equations obtained this way is not unique - it only corresponds to one of the possible sets of equations that have the 2nd order gradient expansion as their asymptotic solution. At this level, this can be viewed as a type of UV completion procedure of the gradient expansion equations that is consistent with the asymptotic causality condition. Alternatively, this general procedure can be illustrated via a simple classical mechanics example. The differential equations $\ddot{x}+\gamma \dot{x}+x = f(t)$ and $\gamma \dot{x}+x=f(t)$ have the same asymptotic solution $x_{asymp}(t) \sim f(t)$ for large times $t \gamma \gg 1$ though their transient (short time $t \gamma \sim 1$) behavior can be very different.}. These equations are similar to those used in current viscous hydrodynamic codes but they include additional 2nd order terms that are usually not taken into account.

This simplified theory still contains 13 transport coefficients and we have presented in this paper either direct calculations or leading estimates for all of them. In the case of the transport coefficients $\tau_{\pi}^*$, $\lambda_1$, $\lambda_2$, $\xi_1$, and $\xi_2$ much more work needs to be done to obtain their exact temperature dependence in our holographic non-conformal model with a crossover phase transition. At the moment, our best estimate for their temperature behavior consisted in using the known expressions for $\lambda_1$ and $\lambda_2$ from SYM and the expressions for $\tau_{\pi}^*$, $\xi_1$, and $\xi_2$ from the (related) non-conformal model of \cite{Kanitscheider:2009as}. This (admittedly uncontrolled) approximation must be taken with care, as emphasized in the main text. However, given our current ignorance regarding these coefficients, we believe that it still may be of phenomenological interest to heavy ion collisions to use these expressions and investigate their consequences. In particular, the direct shear-bulk coupling term $\tau_\pi^*$ may be very relevant in hydrodynamic simulations, as emphasized in \cite{Denicol:2014vaa} and \cite{Denicol:2014mca,Jaiswal:2014isa}.

Also, regarding the complete evaluation of the 2nd order transport coefficients that appear in the gradient expansion of non-conformal strongly-coupled hydrodynamics, the fluid/gravity correspondence \cite{Bhattacharyya:2008jc} provides a way to compute all the coefficients. However, when the background metric is only known numerically, as it is in our case with a crossover phase transition, the actual implementation of the fluid/gravity approach becomes much more challenging. In this context, it may be useful to consider a simpler analytical model that still possesses a phase transition. For instance, one may consider the finite temperature holographic model of Ref.\ \cite{Kajantie:2011nx} where the background metric and scalar field are known analytically and the system displays a 1st-order deconfinement phase transition. In this case, it should be possible to carry out the fluid/gravity procedure and find expressions for all the 2nd order transport coefficients. This would allow for a complete study of entropy production in 2nd order non-conformal hydrodynamics \cite{romatschke} for a theory that displays a phase transition.     

We remark that very similar equations of motion for a non-conformal plasma have been derived from the Boltzmann equation in \cite{Denicol:2014vaa} and, in that paper, the authors also gave explicit formulas for several 2nd order transport coefficients. It would be interesting to compare the results of hydrodynamic simulations computed using the strongly coupled transport coefficients of this paper with those obtained using the kinetic theory-derived coefficients of \cite{Denicol:2014vaa}\footnote{The transport coefficients in \cite{Denicol:2014vaa} were computed using the 14-moment approximation for the relativistic Boltzmann equation. Thus, their results are not valid for a strongly coupled fluid with a crossover phase transition. On the other hand, one should also keep in mind that the holographic approach pursued here is certainly not applicable at low temperatures (where a Boltzmann description of hadron dynamics should be applicable) or at sufficiently high temperatures (where asymptotic freedom is dominant).}. We have gathered in Appendix \ref{summaryhydro} the fitting functions that describe the temperature dependence of all the transport coefficients in this 2nd order Israel-Stewart theory to facilitate their use in current hydrodynamic codes.

This brings us to an important point concerning the equations of motion of strongly-coupled relativistic hydrodynamics in the light of the gauge/gravity duality. As discussed in \cite{Denicol:2011fa,Noronha:2011fi}, the fact that the non-hydrodynamic modes of the $xy-xy$ energy-momentum tensor retarded correlator possess comparable real and imaginary parts at zero momentum \cite{Kovtun:2005ev} implies that, strictly speaking, the effective theory that should be able to describe the hydrodynamical sound and shear modes as well as the lowest set of non-hydrodynamic modes is not of relaxation-type such as in Israel-Stewart theory (as obtained from the Boltzmann equation). Ref.\ \cite{Heller:2014wfa} has recently proposed a way to describe the approach towards hydrodynamics in strongly-coupled SYM that involves equations of motion that are qualitatively different than those in (\ref{definenewshearstress1}) and (\ref{definenewbulk1}) since they involve a second order, homogeneous differential equation for the part of the shear stress tensor associated with the two lowest quasinormal modes $\pi^{\mu\nu}_{QNM}$ (also, in their effective approach nonlinear terms in $\pi^{\mu\nu}_{QNM}$ are not taken into account). It would be interesting to investigate whether their effective theory is also applicable in the case of the non-conformal plasma studied in this paper.

\acknowledgments

This work was supported by Funda\c c\~ao de Amparo \`a Pesquisa do Estado de S\~ao Paulo (FAPESP) and Conselho Nacional de Desenvolvimento Cient\'ifico e Tecnol\'ogico (CNPq). The authors thank R.~Critelli, G.~S.~Denicol, M.~Gyulassy, and J.~Noronha-Hostler for useful discussions.

\appendix
\section{Solution for the metric perturbation up to $\mathcal{O}(\omega^2,q^2)$}
\label{apa}

In this Appendix we formally solve Eq.\ \eqref{2.17} in detail in terms of the coefficients of the undisturbed background metric, $g_{MN}^{(0)}(u)$, and the boundary value of the metric perturbation, $\varphi(\omega,q)$, up to $\mathcal{O}(\omega^2,q^2)$. In order to accomplish that one must specify the boundary conditions for the metric perturbation, $\phi(u,\omega,q)$, at the boundary and at the horizon.

We first consider the boundary. The asymptotic form of Eq.\ \eqref{2.17} near the boundary $u=\epsilon$ for any asymptotically AdS background reads\footnote{For asymptotically AdS geometries one finds $g_{uu}(\epsilon)\sim g_{tt}(\epsilon)\sim g_{xx}(\epsilon)\sim L^2/\epsilon^2$.}
\begin{align}
\phi_\epsilon ''-\frac{3}{u}\phi_\epsilon '-k^2\phi_\epsilon=0\Rightarrow \phi_\epsilon=C_1(k) \epsilon^2 K_2(k\epsilon) + C_2(k) \epsilon^2 I_2(k\epsilon),
\label{a1}
\end{align}
where $k^2=-\omega^2+q^2$ and $I_n(\xi)$ and $K_n(\xi)$ are the modified Bessel functions of the first and second kinds, respectively. Now we take
\begin{align}
\lim_{u\rightarrow 0}\phi(u,k)=\lim_{\epsilon\rightarrow 0}\phi_\epsilon(k)= \frac{2C_1(k)}{k^2},
\label{a2}
\end{align}
which is a constant in the radial direction. Therefore, in the case of a massless scalar field, one can safely impose the Dirichlet boundary condition as follows
\begin{align}
\lim_{u\rightarrow 0}\phi(u,k)=\varphi(k)\,,
\label{a3}
\end{align}
where $\varphi(k)$ denotes the boundary value of the metric perturbation prescribed by the Dirichlet boundary condition.


Now that we have specified the boundary condition \eqref{a3}, let us discuss which kind of condition we must impose on the metric perturbation at the horizon. As discussed in \cite{ss,prescription-justification-1,prescription-justification-2,full-prescription}, in order to obtain the retarded propagator we must single out the solution of the equation of motion (\ref{2.17}) which is regular at the horizon and corresponds to a wave being absorbed by the horizon. We begin by assuming that the $g_{tt}$ component of the background metric has a simple zero at the horizon $u=u_H$ such that one can write
\begin{align}
g_{uu}(u)=\frac{G(u)}{u_H-u},\,\,\,g_{tt}(u)=F(u)(u_H-u),
\label{a4}
\end{align}
with $G(u_H)$ and $F(u_H)$ finite. Notice that the Hawking temperature of the black brane gives
\begin{align}
T=\frac{\sqrt{g'_{tt} g^{uu}\,'}}{4\pi}\biggr|_{u=u_H}=\frac{1}{4\pi}\sqrt{\frac{F(u_H)}{G(u_H)}}\Rightarrow \frac{F(u_H)}{G(u_H)}=(4\pi T)^2\,.
\label{a5}
\end{align}
Using Eqs.\ \eqref{a4} and \eqref{a5}, the asymptotic form of Eq.\ \eqref{2.17} near the horizon $u=u_H$ then reads\footnote{We only keep the dominant terms in each order in radial derivatives of the metric perturbation.}
\begin{align}
\phi_H ''-\frac{1}{u_H-u}\phi_H '+\frac{(\omega/4\pi T)^2}{(u_H-u)^2} \phi_H=0\Rightarrow \phi_H=C_1(\omega) (u_H-u)^{-i\omega/4\pi T} + C_2(\omega) (u_H-u)^{+i\omega/4\pi T},
\label{a6}
\end{align}
and the infalling wave mode at the horizon is obtained by setting $C_2(\omega)=0$ in Eq.\ \eqref{a6}. Therefore, one is motivated to separate the infalling behavior of the solution and take the following Ansatz for the metric perturbation
\begin{align}
\phi(u,k)&=\varphi(k)u_H^{+i\omega/4\pi T}(u_H-u)^{-i\omega/4\pi T}f(u,k),\label{a7}\\
f(u,k)&=f_0(u)+\omega f_1(\omega)+\frac{\omega^2}{2}f_2(u)+\frac{q^2}{2}f_3(u)+\mathcal{O}(\omega q^2,\omega^3)\,,
\label{a8}
\end{align}
with $f(0,k)=1$ and $f(u_H,k)$ being regular. We considered only even powers of $q$ in the series expansion \eqref{a8} due to spatial isotropy. Substituting Eq.\ \eqref{a7} into Eq.\ \eqref{2.17} we obtain the differential equation for $f(u,k)$,
\begin{align}
\partial_u\left[\sqrt{-g^{(0)}}g^{uu}\left(f'+\frac{i\omega f}{4\pi T(u_H-u)}\right)\right] &+\sqrt{-g^{(0)}}g^{uu}\left(f'+\frac{i\omega f}{4\pi T(u_H-u)}\right)\frac{i\omega f}{4\pi T(u_H-u)}=\nonumber\\
&=\sqrt{-g^{(0)}}\left(-g^{tt}\omega^2 +g^{xx}q^2\right)f.
\label{a9}
\end{align}
Now we solve Eq.\ \eqref{a9} order by order in the momentum expansion \eqref{a8}. 

\begin{description}

\item[Zeroth order - $f_0$] \hfill \\
At lowest order, one obtains the differential equation for $f_0(u)$,
\begin{align}
\mathcal{O}(\omega^0,q^0):\,\,\, \partial_u\left(\sqrt{-g^{(0)}}g^{uu}f_0 '\right)=0.
\label{a10}
\end{align}
One can integrate Eq.\ \eqref{a10} twice to obtain
\begin{align}
f_0(u)=c_1+c_2\int_0^u \frac{d\xi}{\sqrt{-g^{(0)}(\xi)}g^{uu}(\xi)},
\label{a11}
\end{align}
where we just performed two indefinite integrations: the lower limit in the integral present in Eq.\ \eqref{a11} has been conveniently chosen to lie at the boundary; however, we can choose any other value of the radial coordinate to be the lower limit of this integral and each possible different choice would just redefine the integration constants $c_1$ and $c_2$. We immediately fix these constants by using the boundary condition $f(0,k)=1$ and the regularity condition for $f(u_H,k)$. The former fixes $c_1=1$ and the latter fixes $c_2=0$; therefore,
\begin{align}
f_0(u)=1\,.
\label{a12}
\end{align}

\item[First order - $f_1$] \hfill \\
After using Eqs.\ \eqref{a8}, \eqref{a9}, and \eqref{a12}, the differential equation for $f_1(u)$ then reads
\begin{align}
\mathcal{O}(\omega,q^0):\,\,\, \partial_u\left[\sqrt{-g^{(0)}}g^{uu}\left(f_1'+\frac{i}{4\pi T(u_H-u)}\right)\right]=0.
\label{a13}
\end{align}
Integrating once, we obtain,
\begin{align}
\sqrt{-g^{(0)}}g^{uu}\left(f_1'+\frac{i}{4\pi T(u_H-u)}\right)=c_2,
\label{a14}
\end{align}
and, then, integrating again, we find
\begin{align}
f_1(u)=c_1+\int_0^u d\xi\left[\frac{c_2}{\sqrt{-g^{(0)}(\xi)}g^{uu}(\xi)}-\frac{i}{4\pi T(u_H-\xi)}\right]\,.
\label{a15}
\end{align}
Applying the boundary condition $f(0,k)=1$ we obtain $c_1=0$. The horizon regularity condition implies
\begin{align}
c_2=\frac{i\sqrt{-g^{(0)}(u_H)}}{4\pi TG(u_H)}=ig_{xx}^{3/2}(u_H)
\label{a16}
\end{align}
and, therefore,
\begin{align}
f_1(u)=i\int_0^u d\xi\left[ \frac{g_{xx}^{3/2}(u_H)}{\sqrt{-g^{(0)}(\xi)}g^{uu}(\xi)}-\frac{1}{4\pi T(u_H-\xi)}\right]\,.
\label{a17}
\end{align}

\item[Second order - $f_2$] \hfill \\
Using Eqs.\ \eqref{a8}, \eqref{a9}, \eqref{a12}, \eqref{a14}, and \eqref{a16} the differential equation for $f_2(u)$ reads
\begin{align}
\mathcal{O}(\omega^2,q^0):\,\,\, \partial_u\left[\sqrt{-g^{(0)}}g^{uu}\left(\frac{f_2'}{2}+\frac{i f_1}{4\pi T(u_H-u)}\right)\right]=\frac{g_{xx}^{3/2}(u_H)}{4\pi T(u_H-u)}-\sqrt{-g^{(0)}}g^{tt}.
\label{a18}
\end{align}
Integrating Eq.\ \eqref{a18} twice and using Eq.\ \eqref{a17}, we obtain
\begin{align}
f_2(u)&=c_1+2\int_0^u \frac{d\lambda}{4\pi T(u_H-\lambda)}\int_0^\lambda d\xi\left[ \frac{g_{xx}^{3/2}(u_H)}{\sqrt{-g^{(0)}(\xi)}g^{uu}(\xi)}-\frac{1}{4\pi T(u_H-\xi)} \right]+\nonumber\\
&+2\int_0^u \frac{d\lambda}{\sqrt{-g^{(0)}(\lambda)}g^{uu}(\lambda)}\left\{ c_2+\int_0^\lambda d\xi \left[ \frac{g_{xx}^{3/2}(u_H)}{4\pi T(u_H-\xi)}-\sqrt{-g^{(0)}(\xi)}g^{tt}(\xi) \right]\right\}.
\label{a19}
\end{align}
In Eq.\ \eqref{a19}, the factors $\left(4\pi T(u_H-\lambda)\right)^{-1}$ and $c_2/\sqrt{-g^{(0)}(\lambda)}g^{uu}(\lambda)$ diverge at the horizon while the factor $\sqrt{-g^{(0)}(\xi)}g^{tt}(\xi)$ diverges at the boundary; therefore, the boundary condition and horizon regularity fix $c_1=0$ and\footnote{Notice that $\sqrt{-g^{(0)}(\lambda)}g^{uu}(\lambda)\biggr|_{\lambda \rightarrow u_H}\sim g_{xx}^{3/2}(u_H)4\pi T(u_H-\lambda)$.}
\begin{align}
c_2&=-g_{xx}^{3/2}(u_H)\int_0^{u_H} d\xi\left[ \frac{g_{xx}^{3/2}(u_H)}{\sqrt{-g^{(0)}(\xi)}g^{uu}(\xi)}-\frac{1}{4\pi T(u_H-\xi)} \right]+\nonumber\\
& - \int_0^{u_H} d\xi \left[ \frac{g_{xx}^{3/2}(u_H)}{4\pi T(u_H-\xi)}-\sqrt{-g^{(0)}(\xi)}g^{tt}(\xi)\right]\,.
\label{a20}
\end{align}
Thus,
\begin{align}
f_2(u)&=2\left\{\int_0^u \frac{d\lambda}{4\pi T(u_H-\lambda)}\int_0^\lambda d\xi\left[ \frac{g_{xx}^{3/2}(u_H)}{\sqrt{-g^{(0)}(\xi)}g^{uu}(\xi)}-\frac{1}{4\pi T(u_H-\xi)} \right]\right.+\nonumber\\
&+\int_0^u \frac{d\lambda}{\sqrt{-g^{(0)}(\lambda)}g^{uu}(\lambda)}\left(g_{xx}^{3/2}(u_H)\int_{u_H}^0 d\xi \left[ \frac{g_{xx}^{3/2}(u_H)}{\sqrt{-g^{(0)}(\xi)}g^{uu}(\xi)}-\frac{1}{4\pi T(u_H-\xi)} \right]+\right.\nonumber\\
&\left.\left.+\int_{u_H}^\lambda d\xi
\left[\frac{g_{xx}^{3/2}(u_H)}{4\pi T(u_H-\xi)}-\sqrt{-g^{(0)}(\xi)}g^{tt}(\xi) \right]\right)\right\}.
\label{a21}
\end{align}

\item[Second order - $f_3$] \hfill \\
Using Eqs.\ \eqref{a8}), \eqref{a9}), and \eqref{a12} the differential equation for $f_3(u)$ reads
\begin{align}
\mathcal{O}(\omega^0,q^2):\,\,\, \partial_u\left(\sqrt{-g^{(0)}}g^{uu}f_3 '\right)=2\sqrt{-g^{(0)}}g^{xx}.
\label{a22}
\end{align}
Integrating twice, we find
\begin{align}
f_3(u)=c_1+c_2\int_0^u \frac{d\lambda}{\sqrt{-g^{(0)}(\lambda)}g^{uu}(\lambda)} +2\int_0^u \frac{d\lambda}{\sqrt{-g^{(0)}(\lambda)}g^{uu}(\lambda)}\int_0^\lambda d\xi \sqrt{-g^{(0)}(\xi)}g^{xx}(\xi).
\label{a23}
\end{align}
The boundary condition fixes $c_1=0$ and the horizon regularity condition fixes
\begin{align}
c_2=-2\int_0^{u_H} d\xi \sqrt{-g^{(0)}(\xi)}g^{xx}(\xi).
\label{a24}
\end{align}
Consequently,
\begin{align}
f_3(u)=2\int_0^u \frac{d\lambda}{\sqrt{-g^{(0)}(\lambda)}g^{uu}(\lambda)} \int_{u_H}^\lambda d\xi \sqrt{-g^{(0)}(\xi)}g^{xx}(\xi)\,.
\label{a25}
\end{align}

\section{Summary of the transport coefficients for hydrodynamic simulations}\label{summaryhydro}

The equations of motion for the Israel-Stewart-like theory in flat spacetime are (\ref{definenewshearstress1}) and (\ref{definenewbulk1}) together with the conservation equations (\ref{conservationeqs}). The 13 transport coefficients in this theory, namely $\eta/s$, $\zeta/s$, $\tau_\pi$, $\tau_\Pi$, $\tau_\pi^{*}$, $\lambda_1$, $\lambda_2$, $\lambda_3$, $\lambda_4$, $\xi_1$, $\xi_2$, $\xi_3$, and $\xi_4$ were discussed already in the main text but here we gather the formulas that describe them in a single place with the intention to facilitate their use in hydrodynamic simulations.

Besides $\eta/s=1/(4\pi)$, the shear relaxation coefficient $\tau_\pi \eta/T^2$ is described by the function in (\ref{eq:kappafit}) with parameters in Table \ref{tab:parameterstaupi}. $\zeta/s$ is described by the fitting function in Eq.\ (\ref{eq:zetafit}) with the parameters in Table \ref{tab:parameterszeta}. The lower bound for the bulk relaxation coefficient $\tau_\Pi T$ is described by (\ref{eq:tauPIfit}) with parameters in Table \ref{tab:parameterstauPI}. The coefficients $\lambda_3=-\lambda_4=2\kappa^*$, where $\kappa^*$ in (\ref{2.2}) is defined in terms of the coefficient $\kappa$. One can find a fit for $\kappa/T^2$ in (\ref{eq:kappafit}) with parameters in Table \ref{tab:parameterskappa}. The coefficients $\lambda_1$ and $\lambda_2$ can be found in Eq.\ (\ref{lambdaSYM}). Moreover, the coefficients $\xi_1$, $\xi_2$, and $\tau_\pi^*$ are found in Eqs.\ (\ref{x1x2}) and (\ref{taupiestrela}), respectively. The final coefficients $\xi_3$ and $\xi_4$ are given by Eqs.\ (\ref{eq:xi32}) and (\ref{eq:xi42}) and can be computed using the results for the previous coefficients discussed above. Furthermore, in all of these fitting functions and tables, the parameter $T_c$ is equal to 143.8 MeV.  

Given that the thermodynamic properties of the model are very similar to those found on the lattice \cite{fodor2012} when $T\sim 130 - 450$ MeV, if one wants to use the transport coefficients computed in this paper in hydrodynamic simulations of the QGP formed in heavy ion collisions one may just use directly an interpolation for the lattice data when computing $c_s^2$ and the entropy density $s$ needed in the evaluation of the transport coefficients. 

\section{Linear instability of the gradient expansion at 2nd order}\label{appendixhugo}

In this Appendix we investigate the {\it linear stability} properties of a fluid described by the 2nd order gradient expansion theory in \eqref{defineshearstress} and \eqref{definebulk} (in flat spacetime) around static equilibrium. In a linear analysis, the relevant linear terms involving the dissipative part of the energy-momentum tensor are
\begin{eqnarray}
\pi^{\mu \nu} = - \eta \sigma^{\mu \nu} + \eta \tau_{\pi} D \sigma^{<\mu \nu>} \,\,{,} \nonumber \\
\Pi = -\zeta \theta + \zeta \tau_{\Pi} D \theta \,\,{.}
\end{eqnarray}
We follow \cite{Pu:2009fj,acausal,shipu1,Romatschke:2009im} and consider linear perturbations around a static background. In order to investigate the stability of the sound channel, it is sufficient to study the effect of the perturbations
\begin{eqnarray}
\varepsilon = \varepsilon_{0} + \delta \varepsilon(t,x)  \,\,{,} \nonumber \\
P = P_{0} + \delta P(t,x)  \,\,{,} \nonumber \\
u_{sound}^{\mu} = (1,0,0,0) + (0,\delta u^{x} (t,x),0,0) \,\,{,} \nonumber \\ 
\eta = \eta_{0} + \delta \eta(t,x) \,\,{,} \, \, \, \tau_{\pi} = \tau_{\pi, 0} + \delta \tau_{\pi}(t,x) \,\,{,} \nonumber \\ 
\zeta = \zeta_{0} + \delta \zeta(t,x) \,\,{,} \, \, \, \tau_{\Pi} = \tau_{\Pi, 0} + \delta \tau_{\Pi}(t,x) \,\,{.} 
\end{eqnarray}
In this case, the relevant terms for linear perturbations are
\begin{eqnarray}
\theta = \partial_{x} \delta u^{x} \,\,{,} \nonumber \\
\sigma^{x x} = \frac{4}{3} \partial_{x} \delta u^{x} + \mathcal{O}(\delta^{2}) \,\,{,} \nonumber \\
\pi^{x x} = - \frac{4}{3} \eta_{0} \partial_{x} \delta u^{x} + \frac{4}{3} \eta_{0} \tau_{\pi, 0} \partial_{x} \partial_{t} \delta u^{x} + \mathcal{O}(\delta^{2}) \,\,{,} \nonumber \\
\Pi = - \zeta_{0} \partial_{x} \delta u^{x}  + \zeta_{0} \tau_{\Pi,0} \partial_{t} \partial_{x} \delta u^{x} + \mathcal{O}(\delta^{2}) \,\,{.}
\label{perturbations1}
\end{eqnarray}
Using these results in the conservation equations \eqref{conservationeqs}, one obtains the following differential equation for the sound disturbance\footnote{Note that we use dimensionless variables $t \to t \,T_0$ and $x \to x T_0$ and, correspondingly, $\omega \to \omega/ T_0$ and $k \to k/ T_0$.}
\begin{equation}
\left[ \partial_{t}^{2} - c_{s,0}^{2} \partial_{x}^{2} - \left( \frac{4}{3} \frac{\eta_{0}}{s_{0}} + \frac{\zeta_{0}}{s_{0}} \right) \partial_{x}^{2} \partial_{t} +  \left( \frac{4}{3} \frac{\eta_{0}}{s_{0}} \tau_{\pi,0} + \frac{\zeta_{0}}{s_{0}} \tau_{\Pi,0}  \right) \partial_{x}^{2} \partial_{t}^{2}  \right] \delta u^{x}(t,x) = 0 \,\,{,}
\end{equation}
where $s_0 T_0 = \varepsilon_0+P_0$. In Fourier space, for $\delta u^{x}(t,x) =  \delta u_{0}^{x} \, e^{i (k x - \omega t)} $, one finds the dispersion relation 
\begin{equation}
\omega^{2} - c_{s,0}^{2} k^{2} +  \left( \frac{4}{3} \frac{\eta_{0}}{s_{0}} + \frac{\zeta_{0}}{s_{0}} \right) i \omega k^{2} - \left( \frac{4}{3} \frac{\eta_{0}}{s_{0}} \tau_{\pi,0} + \frac{\zeta_{0}}{s_{0}} \tau_{\Pi,0}  \right) \omega^{2} k^{2} = 0  \,\,{.}
\end{equation}
While the equation above can be solved exactly, when it comes to the stability properties of these modes it is sufficient to look at the sum of the roots \cite{hugofuturo}. For the polynomial corresponding to sound disturbances, the sum of the two roots gives
\begin{eqnarray}
\omega_{1} + \omega_{2} &=& \frac{i\,\left( \frac{4}{3} \frac{\eta_{0}}{s_{0}} + \frac{\zeta_{0}}{s_{0}} \right)}{\left( \frac{4}{3} \frac{\eta_{0}}{s_{0}} \tau_{\pi,0} + \frac{\zeta_{0}}{s_{0}} \tau_{\Pi,0}  \right) k^{2} - 1}  \,\,{.}  
\end{eqnarray}
Notice that for $k$ larger than a critical wavenumber $k_{c}^{sound}$ defined by
\be
k_{c}^{sound} = \frac{1}{\sqrt{ \frac{4}{3} \frac{\eta_{0}}{s_{0}} \tau_{\pi,0} + \frac{\zeta_{0}}{s_{0}} \tau_{\Pi,0}} } \,\,{,}
\ee
the sum of the roots adds up to a positive imaginary number. Therefore, for $k > k_c^{sound}$ one of the modes has a positive imaginary part and is, thus, unstable. 

Note that in the limit when $\tau_{\pi,0}\,,\tau_{\Pi,0} \to 0$ one finds that $k_c^{sound}$ diverges, which is in agreement with the fact that NS theory is stable against small perturbations in a fluid at rest \cite{acausal}. Clearly, for a moving fluid the stability properties become more involved but it is possible to show that the same type of problems that appears in NS theory also appear in this case \cite{hugofuturo}. However, we remark that hydrodynamics is only expected to be valid in the low frequency, large wavelength limit. Nevertheless, the instability found here at finite wavenumber motivates the search for a UV completion of this theory (for instance, the one discussed in Section \ref{ISsection}) that is linearly stable and can, therefore, be safely used in numerical simulations.

A larger critical wavenumber, $k_c^{shear}$, appears in the shear channel. Linear stability of shear modes can be studied by choosing a flow disturbance of the kind $u^{\mu}_{shear} = (1,0,0,0) + (0,0, \delta u^{y}(t,x),0 )$ while the other relations in \eqref{perturbations1} remain valid (see \cite{Romatschke:2009im}). This leads to the following dispersion relation
\begin{equation}
\omega \left( 1 - \frac{\eta_{0}}{s_{0}} \tau_{\pi_{0}} k^{2} \right) + i \frac{\eta_{0}}{s_{0}} k^{2} = 0 \,\,{,}
\end{equation}
which can be easily solved to give
\begin{eqnarray}
\omega(k) = \frac{i\, \frac{\eta_{0}}{s_{0}} k^{2}}{  \frac{\eta_{0}}{s_{0}} \tau_{\pi,0} k^{2} - 1} \,\,{,} \nonumber \\ 
k_{c}^{shear} = \frac{1}{\sqrt{\frac{\eta_{0}}{s_{0}} \tau_{\pi,0}}}>k_c^{sound} \,\,{.} 
\end{eqnarray}
For a recent study involving the first nonlinear corrections to the stability analysis and the propagation of waves in relativistic hydrodynamics see, for instance, \cite{Fogaca:2014gwa}.

\end{description}


\end{document}